\documentclass[a4paper,fleqn]{cas-sc}

\usepackage[numbers]{natbib}
\usepackage{graphicx}
\usepackage{subcaption}
\usepackage[utf8]{inputenc}
\usepackage[export]{adjustbox}
\usepackage{wrapfig}
\usepackage{blindtext}

\usepackage{amssymb}
\usepackage{gensymb}

\usepackage{subcaption}

\usepackage{afterpage}

\usepackage{framed} % Framing content
\usepackage{multicol} % Multiple columns environment
\usepackage{nomencl} % Nomenclature package

\makenomenclature
\renewcommand*\nompreamble{\begin{multicols}{2}}
\renewcommand*\nompostamble{\end{multicols}}

\usepackage{etoolbox}
\renewcommand\nomgroup[1]{%
  \item[\bfseries
  \ifstrequal{#1}{A}{Alphabetic}{%
  \ifstrequal{#1}{S}{Subscripts}{%
  \ifstrequal{#1}{O}{Other Symbols}{%
  \ifstrequal{#1}{B}{Abbreviation}{}}}}%
]}

\def\tsc#1{\csdef{#1}{\textsc{\lowercase{#1}}\xspace}}
\tsc{WGM}
\tsc{QE}
\tsc{EP}
\tsc{PMS}
\tsc{BEC}
\tsc{DE}
\begin{document}
\let\WriteBookmarks\relax
\def\floatpagepagefraction{1}
\def\textpagefraction{.001}
\shorttitle{Boiling on vertical cylinders}
\shortauthors{A. Nikulin et~al.}
%\begin{frontmatter}

\title [mode = title]{Spacing effect on pool boiling performance of three triangular pitched and vertically oriented tubes}                      
%\tnotemark[1,2]

%\tnotetext[1]{This document is the results of the research
%   project funded by the National Science Foundation.}

%\tnotetext[2]{The second title footnote which is a longer text matter
%   to fill through the whole text width and overflow into
%   another line in the footnotes area of the first page.}

\author[1]{Artem Nikulin\corref{cor1}}[orcid=0000-0002-3304-0506]\cormark[1]

\author[1]{Jean-Luc Dauvergne}
\author[1]{Asier Ortuondo}
\author[1,2]{Elena Palomo del Barrio}

\address[1]{Centre for Cooperative Research on Alternative Energies (CIC energiGUNE), Basque Research and Technology Alliance (BRTA), Alava Technology Park, Albert Einstein 48, 01510 Vitoria-Gasteiz, Spain}
\address[2]{Ikerbasque --- Basque Foundation for Science, Mar\'ia D\'iaz Haroko 3, 48013 Bilbao, Spain}

%\cortext[cor1]{Corresponding author: anikulin@cicenergigune.com}

\nonumnote{*Corresponding author: anikulin@cicenergigune.com}

\begin{abstract}
There is a scarcity of available data on boiling process in vertically oriented tube bundles in accessible sources. Lack of systematic studies is limiting further expansion of this highly efficient process of heat transfer into heat recovery field. In this paper boiling process of three triangular pitched and vertically oriented tubes has been studied in ethanol at 78$^{\circ}$C. The main focus of this work was to study the effect of tube spacings on heat transfer coefficient (HTC) and bubbles behavior (bubble departure diameter in particular) that were visualised with the help of a high speed camera. Experiments were performed in a wide range of tube spacings (from 10.75 to 0.25 mm) and heat flux densities (from 3 to 70 kW/m$^2$). 

The obtained results show that, long spacings i.e., much longer than bubble departure diameter, have no influence on HTC as well as on bubbles behavior. On the contrary, the spacings on the order of the bubble departure diameter tend to create slug flow in the bundle, that is very beneficial for the heat exchange at low heat fluxes. Finally, narrow spacings that are much shorter than the bubble departure diameter have shown the potential to enhance the HTC in tube bundles with low length to diameter ratios.
 \end{abstract}

%\begin{graphicalabstract}
%\includegraphics{figs/grabs.pdf}
%\end{graphicalabstract}

%\begin{highlights}
  %  \item Long spacings have no influence on heat transfer in tube bundles. 
  %  \item Slug flow in a bundle is very beneficial for heat exchange at low heat fluxes
 %   \item Spacings shorter than bubble departure diameter enhance heat exchange at low length to diameter ratios
    
%\end{highlights}


\maketitle

\section{Introduction}
\par
Liquid-vapor phase transition is a basis of multiple industrial processes. For instance, it is vital for vapor-compression refrigeration systems such as air conditioners, refrigerators, heat pumps and chillers. It is widely used in nuclear and thermal power plants, chemical, food and various others industries. Usually it accrues in the evaporators or boilers of such equipment through the evaporation and can be accompanied with boiling. 

As well known, pool boiling \cite{nikulin2018study} provides much higher heat transfer coefficients (HTC) compared to convective \cite{nikulin2019effect} and heat transfer under evaporation \cite{weibel2010characterization,parken1990heat} at similar conditions. Furthermore, application of pool boiling allows to create economically more feasible and technically more reliable systems that not require an external pump and extra heat exchanger as compared to flow boiling \cite{hayes2019regulating}. Consequently, the effect of tubes arrangement on HTC it important information for effective heat exchangers design in a sense of their size and material consumption.

In spite of a significant amount of studies focused on the boiling on tubes and tubes bundles in a confined and unconfined conditions, this process is not fully explored. As follow from the reviews \cite{ren2020pre,swain2014review}, most of the published papers are dealing with horizontal tubes bundles, rather than with vertically oriented. Nonetheless, it has been shown, that vertically aligned single tube \cite{kang2000effect,chun1998effects}, confined single tube \cite{kang2010pool} and tubes bundles in a heat exchanger \cite{zhang2017experimental} may have higher HTC compared to horizontal arrangement under similar conditions. 

Gupta et al. \cite{gupta2010nucleate} studied boiling process of saturated water on a bundle of seven stainless steel (SS) vertical tubes having 19 mm in diameter (D), pitch to diameter ratio (P/D) of 1.66 and length to diameter ratio (L/D) of 45. They found, that the HTC linearly increases with height of the tubes in the whole range of studied heat fluxes from 2 to 32 kW/m$^2$. It was attributed to the higher turbulence produced by floating up bubbles. Moreover, the onset of boiling happened in the lower part of the bundle. In the study of Chung et al. \cite{chung2015heat} the boiling performance of vertical bundle of seven SS tubes having D of 21.3 mm, P/D of 1.53 and L/D of 70 was tested with sub-cooled water as working fluid. The authors had reported similar to Gupta et al. findings \cite{gupta2010nucleate} i.e. the HTC increases along the height of the tubes bundle due to higher turbulence caused by rising bubbles and the onset of sub-cooled boiling starts at the lower part of the bundle. In addition, the average HTC of the bundle is slightly higher than the HTC of a single tube, mainly due to the higher HTC of the central tube in the bundle.

To the best of our knowledge, only one study \cite{kang2016effect} is dealing with a pitch effect of two vertical tubes on pool boiling of saturated water. Two P/D of 1.5 and 5.0 were reported for the SS tubes having L/D of 21. The heat flux density of the left-handed tube was set to 0, 30, 60, and 90 kW/m$^2$ while the heat flux density of the other tube varied from 10 to 120 kW/m$^2$. At the shortest P/D of 1.5 the left tube was influencing the HTC of the right one until the heat flux density reaches 40 kW/m$^2$. The more the heat flux set on the left tube the higher impact was registered. The HTC becomes about 1.45 times higher at 90 kW/m$^2$ supplied to the left tube and 10 kW/m$^2$ to the right tube. However, the influence of the left tube on the right one is negligible when P/D is greater than 4. The enhancement of HTC was associated with the increase of the liquid agitation generated by the moving bubbles, that improve convective heat transfer.

The authors interest to the pool boiling in vertically oriented tube bundles is associated with the potential applicability of this process in mobile thermal energy storage (M-TES) systems. M-TES systems have been proposed in order to connect a waste heat supplier with a consumer when pipe-line connection is economically not feasible \cite{miro2016thermal,alva2018overview}. Moreover, it has been shown, that such systems can be more efficient and feasible by means of phase change materials (PCMs) to store thermal energy \cite{deckert2014economic,yabuki2007non,kaizawa2008thermal} even when utilizing low or medium potential waste heat. Considering that major part of the heat is wasted at low and medium potentials \cite{panayiotou2017preliminary,firth2019quantification,bianchi2019estimating} that makes reasonable further development of M-TES systems. According to the study of Deckert et al. \cite{deckert2014economic} not only the storage capacity, but also charging and discharging energy rates play an important role in the economic efficiency of such systems. Thus, more efficient methods to charge and discharge M-TES system, such as boiling, that may significantly increase the number of operating cycles per year have to be explored.

Thus, the aim of this study is to get a set of experimental data regarding the effect of spacing on pool boiling in a wide range of P/D in a vertically oriented tube bundles, that in M-TES applications can represent the storage material inside the storage vessel. Taking into account that the behaviour of bubbles is strongly affecting the HTC while boiling \cite{nikulin2018study,pioro2004nucleate,gerardi2011infrared,pontes2020effect}, the bundle of three triangular pitched tubes was tested in order to compromise the complexity of the phenomenon and the possibility of optical study through video recording. Subsequently, those data will be used to verify CFD modelling of boiling heat transfer and dynamics of bubbles in similar tube bundle configuration with the final aim to simulate and optimise heat storage capacity and discharge rate of M-TES systems for low and medium potential waste heat recovery.

\begin{table}  

\begin{framed}

\nomenclature[A]{$D$}{Tube diameter, [mm]}
\nomenclature[A]{$P$}{Tubes pitch, [mm]}
\nomenclature[A]{$L$}{Tube length, [mm]}
\nomenclature[A]{$S$}{Tubes spacing, [mm]}
\nomenclature[A]{$Q$}{Heat flux, [W]}
\nomenclature[A]{$q$}{Heat flux density, [W/m$^2$]}
\nomenclature[A]{$U$}{Voltage, [V]}
\nomenclature[A]{$I$}{Current, [A]}
\nomenclature[A]{$h$}{Heat transfer coefficient, [W/m$^2\cdot$K]}
\nomenclature[A]{$\overline{h}$}{Average heat transfer coefficient, [W/m$^2\cdot$K]}
\nomenclature[A]{$A$}{Heater surface area, [m$^2$]}
\nomenclature[A]{$H$}{Height of tube, [mm]}
\nomenclature[A]{$D_b$}{Bubble diameter, [mm]}

\nomenclature[A]{$R_a$}{Arithmetical mean roughness [$\mu$m]}
\nomenclature[A]{$R_z$}{Peak to valley roughness, [$\mu$m]}

\nomenclature[B]{$SS$}{Stainless steel}
\nomenclature[B]{$HTC$}{Heat transfer coefficient}

\nomenclature[O]{$\Delta T$}{Wall superheat, [K]}

\printnomenclature
\end{framed}

\end{table}

\section{Experimental section}
\subsection{Experimental setup description and procedure}

\begin{figure}
\centering
\includegraphics[width=1.00\linewidth]{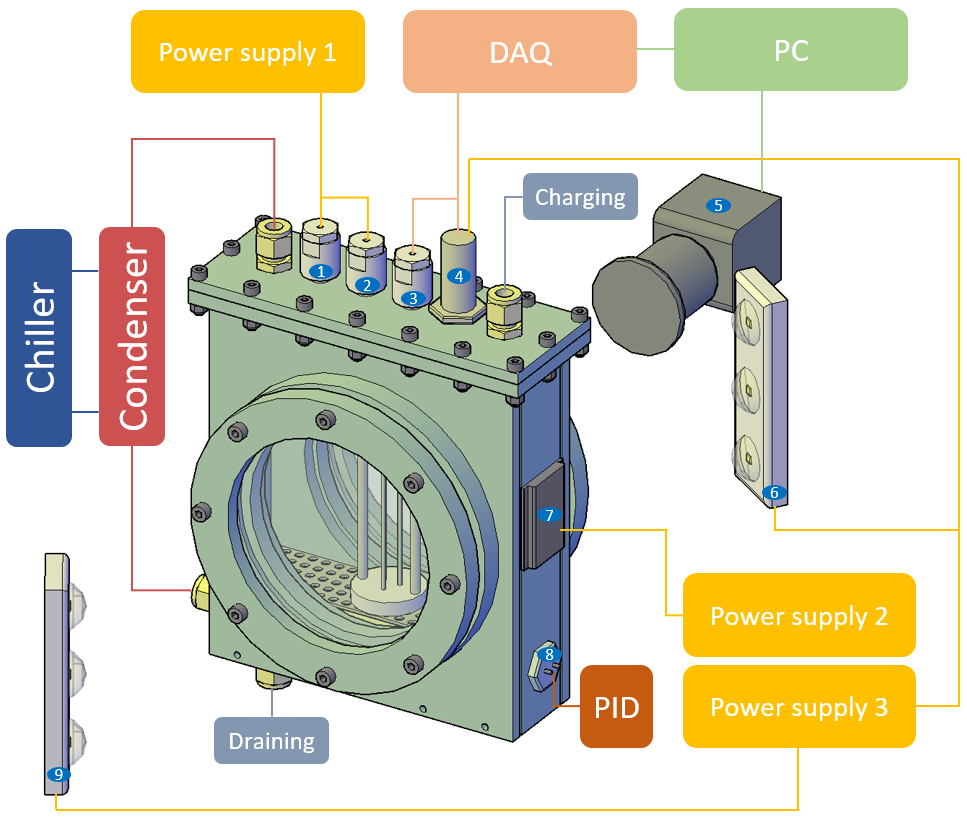}%
\caption{Scheme of the experimental setup: (1,2) power feedthroughs, (3) sensors feedthrough, (4) pressure transducer, (5) high speed camera, (6,9) LEDs, (7) plane heater, (8) cartridge heater.}%
\label{fig:setup}
\end{figure}

A scheme of the experimental setup is shown in Fig. \ref{fig:setup}. It is a closed thermosyphon-like system that consists of a condenser, a Julabo DD-1000FF chiller and a boiler. The main part of the setup is the boiler that is made out of 10 mm stainless steel plates (AISI 304). The width, depth and height of the internal volume of the boiler are 210, 50 and 250 mm respectively. Two plane-parallel sight glasses with 150 mm in diameter viewports are directly welded to the boiler.

The electric power is supplied to the test section from the power supply 1 (EA-PSI 9080-40DT) through two feedthroughs (1,2). The feedthrough (3) is used to connect potential and k-type thermocouples wires supplied from Omega. The setup is also equipped with a PX2 series pressure transducer (4) from Honeywell. All measurements are performed with a DAQ Rigol M300 connected together with an IX-cameras i-SPEED 210 high speed camera at 800 fps (5) to a common computer. The four channels power supply 3 (MX100Q from Aim-TTi) is feeding a pressure transducer (4) and two LEDs assemblies (6,9) installed from each side of the boiler for pool boiling process illumination. A 500W cartridge heater (8) and a PID system are used for fast heating up the boiler to a desired temperature and for deaeration of the working fluid. All parts of the setup exposed to the air, except viewports, are insulated with a 10 mm thick rubber insulation having thermal conductivity less than 0.04 W/(m$\cdot$K). However, in order to fully compensate the heat losses to the ambient from the exterior surface of the vessel, two 110 W plane heaters (7) fed from the power supply 2 (EA-PS 9080-40T) are attached to both sides of the boiler.

Ethanol was used as working fluid. The use of ethanol for heat recovery systems was previously discussed \cite{meinel2014effect,mastrullo2018flow} with a positive conclusion relatively to organic Rankine cycle. Taking into account its high latent heat, relatively low saturation pressure, low degree of superheating \cite{mastrullo2018flow}, environmental friendliness and renewability, ethanol can be considered as promising working fluid for M-TES systems working for low or medium temperature waste heat recovery. Flammability and toxicity of ethanol can be overcome by creation of sealed systems \cite{kalani2013enhanced}.

Before the experiment, ethanol was deaerated by continuous boiling during at least 20 minutes while the vapor/air mixture was exhausted from the upper part of the setup. Afterwards, the system was closed and PID controller connected to the cartridge heater (8) was switched off. Then, the power supplied to plane heaters (7) was adjusted that way that the temperature change rate in the boiler was less than 0.1 K per hour. In all cases it was equal to 70$\pm$5 W. After that, experiment begun. Each data point was obtained at steady state, descending heat flux and desired boiling temperature that was controlled within $\pm$0.5 K by varying  temperature of the liquid pumped through the condenser from the chiller.

Heat flux $Q$ dissipated at the test section was calculated using the following equation

\begin{equation}\label{Q}
 Q = U \cdot I,
 \end{equation}
where, $U$ is the voltage drop across the test section; $I$ is the current in the circuit. 

The local heat transfer coefficient $h$ along the tubes was calculated as follow

\begin{equation}\label{h}
 h = \frac{Q}{A \cdot \Delta T},
 \end{equation}
where, $A$ is the surface area of the heater, m$^2$; $\Delta T$ is temperature difference between local temperature of the heater and boiling temperature, K. 

The maximum temperature difference across the SS capillaries (AISI 321) wall was calculated with the aid of CFD modeling for the worst possible conditions. Thus, at heat flux density of 70 kW/m$^2$, heat transfer coefficient of 7800 W/m$^2\cdot$K and thermal conductivity of SS 16.3 W/m$\cdot$K the temperature differences across the capillaries wall is equal to 0.059 K. Thereby, it was neglected and the local temperatures of the heater surface were assumed to be equal to the temperature on the inner surface measured by thermocouples.

%The time average values of HTC $\overline{h}$ were obtained by numerical integration of local once based on trapezoidal rule that finally were averaged between values obtained for each of three tubes.%

\subsection{Test section configuration and characterization}

\begin{figure}
\begin{center}
\begin{minipage}{0.15\linewidth}
\center{\includegraphics[width=1\linewidth]{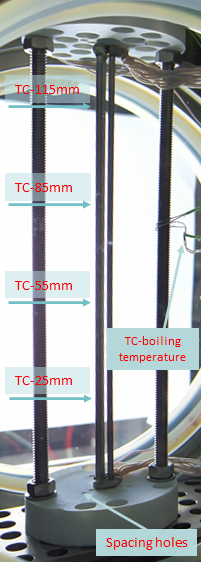}} a)\\
\end{minipage}
\begin{minipage}{0.21\linewidth}
\center{\includegraphics[width=1\linewidth]{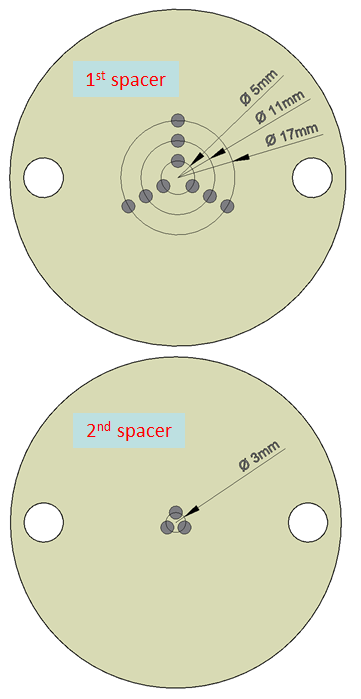}} b)\\
\end{minipage}
\caption{Test section configuration: a) location of thermocouples; b) spacers dimensions}
\label{fig:test sec}
\end{center}
\end{figure}

The test section was made out of three SS capillaries (AISI 321), having 2 mm in outer diameter, 0.1 mm in wall thickness, 130 mm in length and a L/D ratio of 65. Four k-type thermocouples were installed in each capillary  at 25, 55, 85, and 115 mm counted from the bottom, to measure the local temperature distribution along their heights. One more thermocouple was measuring boiling temperature (see Fig. \ref{fig:test sec}(a) for details). The tubes were connected in series to the power supply in order to promote Joule heating and to achieve constant heat flux conditions. Two pairs of spacers were made in order to assure correct positioning of the capillaries. The spacers were PTFE discs with holes of 2 mm. The cross views of the utilized spacers are shown in Fig. \ref{fig:test sec}(b). Four configurations were tested in this work, the corresponding tubes spacing $S$, center to center pitch $P$ and pitch to diameter ratio $P/D$ are listed in Table \ref{tabular:spacing}. After installation of the capillaries the spacers were fixed with the help of two SS studs onto a SS perforated plate inside the boiler and flooded 20 mm above the highest level with 96 \%  ethanol supplied by Scharlau. 

\begin{table}
\caption{Tubes spacing $S$, center to center pitch $P$ and pitch to diameter ratio $P/D$}
\label{tabular:spacing}
\begin{center}
\begin{tabular}{lllll}
\hline
Spacing number  & 1     &     2 &    3 &   4 \\ 
\hline
$S$, mm         & 10.75  &  6.25 & 1.75 & 0.25  \\
$P$, mm         & 12.75  &  8.25 & 3.75 & 2.25  \\
$P/D$           & 6.38  &  4.13 & 1.88 & 1.13    \\
\hline
\end{tabular}
\end{center}
\end{table}

\begin{figure}
\begin{minipage}[h]{0.15\linewidth}
\center{\includegraphics[width=1\linewidth]{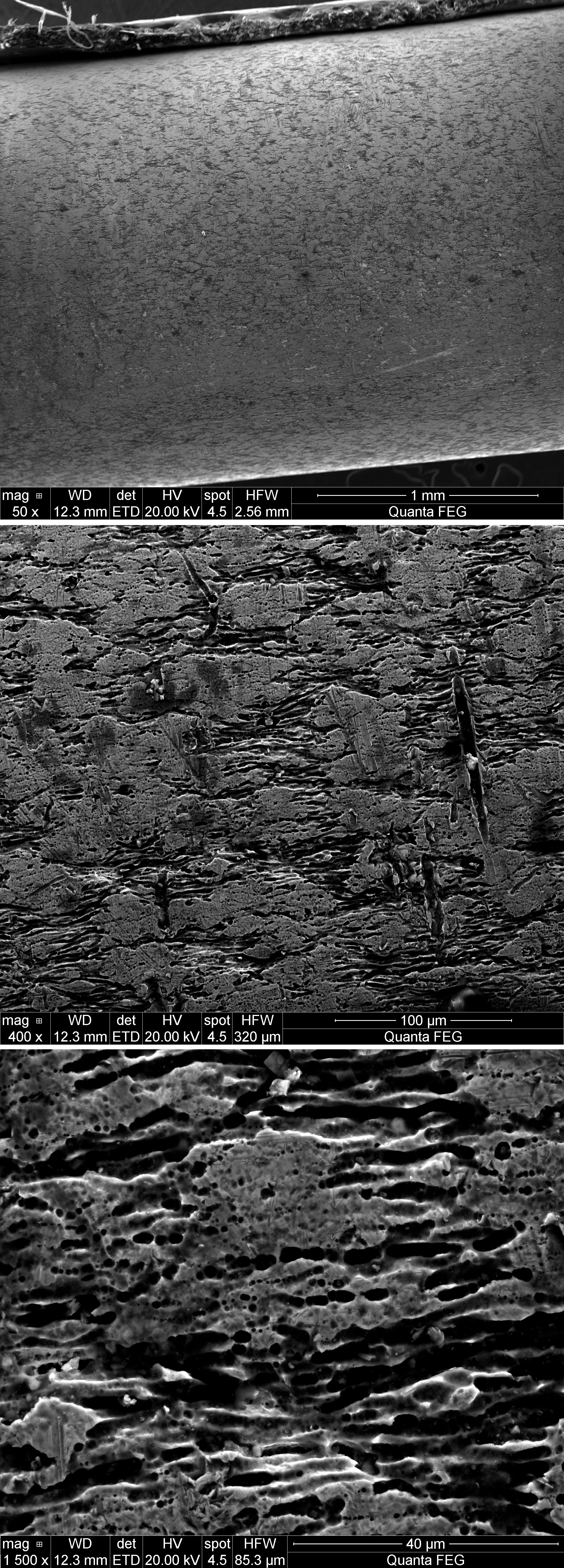}} a) \\
\end{minipage}
\begin{minipage}[h]{0.15\linewidth}
\center{\includegraphics[width=1\linewidth]{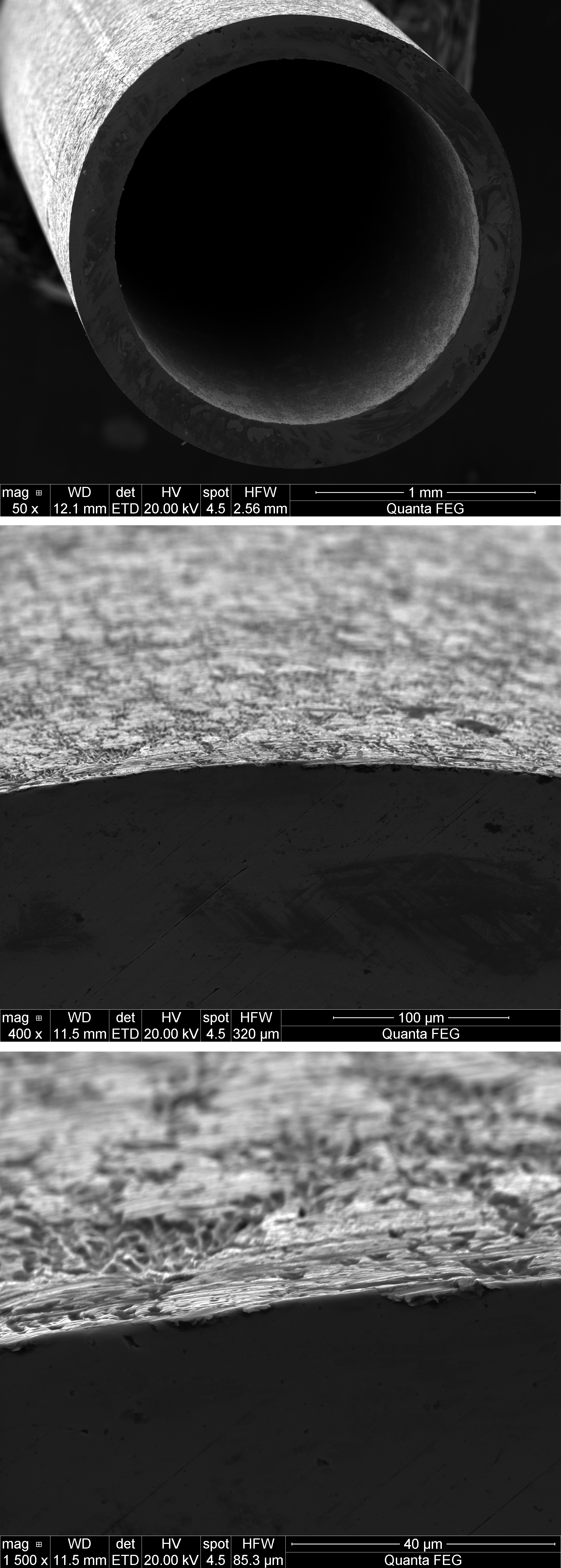}} b) \\
\end{minipage}
\begin{minipage}[h]{0.55\linewidth}
\center{\includegraphics[width=1\linewidth]{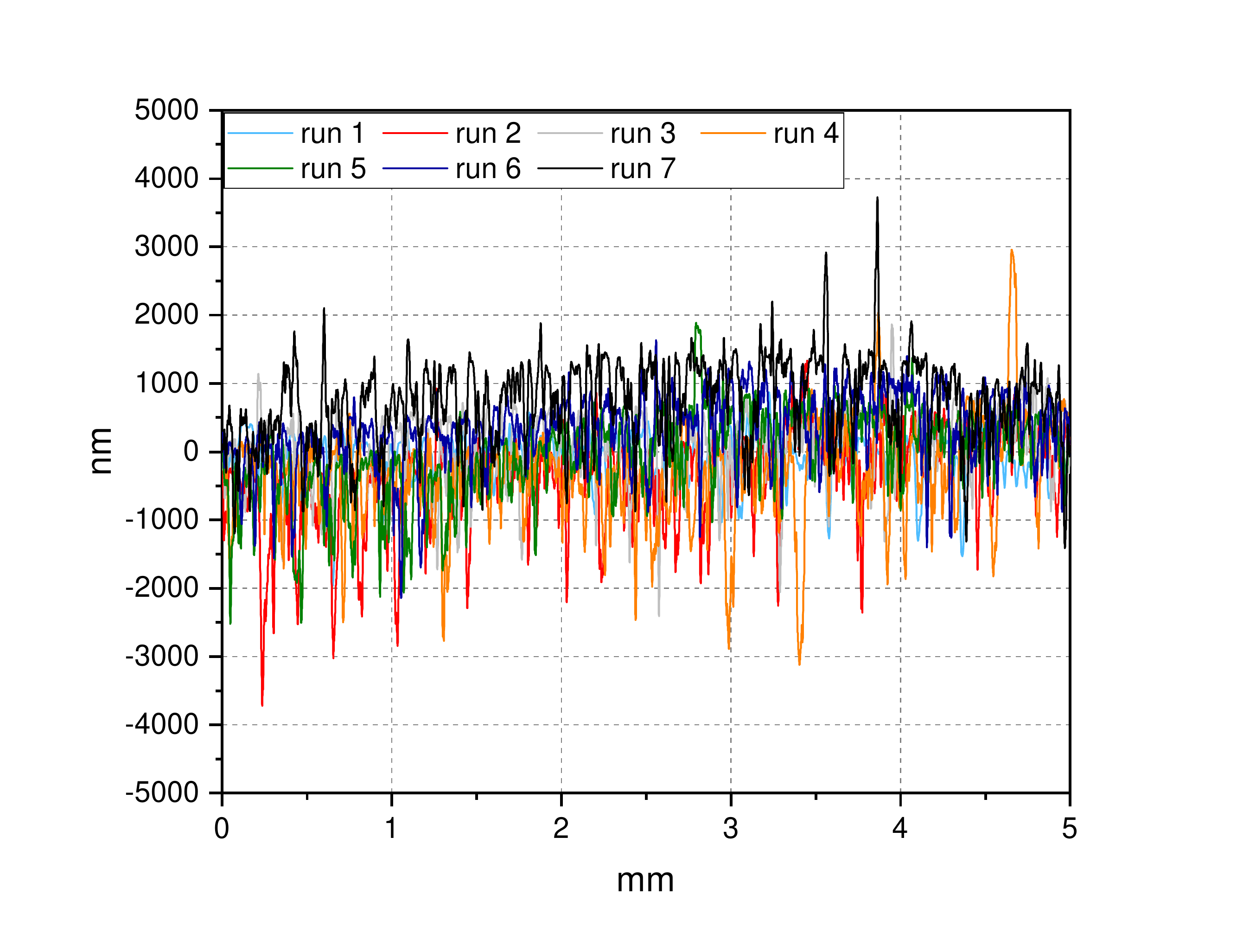}} c) \\
\end{minipage}
\caption{Surface of the SS capillaries: a) SEM images from the top; b) SEM images of the cross section; c) Surface profiles}
\label{fig:surface}
\end{figure}

SS capillaries have been used as received, i.e. the capillaries surface was not treated anyhow (polished, sanded, etched etc.). Their surface was covered by several types of irregularities that are typical for the surfaces produced by drawing method (see SEM images in Fig. \ref{fig:surface}(a) and (b)). The roughness of the external tube surface was analysed by DektakXT (Bruker) stylus profilometer. Seven profiles of the surface taken randomly along the length and diameter as well as the results of arithmetical mean deviations of the measured profiles and maximum peak to valley heights calculations are shown in Fig. \ref{fig:surface}(c) and Table \ref{tabular:roughness}.

\begin{table}
\caption{Results of the roughness calculation}
\label{tabular:roughness}
\begin{center}
\begin{tabular}{lll}
\hline
Run & R$_a$, $\mu$m & R$_z$, $\mu$m \\ 
\hline
run 1 & 0.26  &  2.83  \\
run 2 & 0.49  &  5.06 \\
run 3 & 0.40  &  4.27 \\
run 4 & 0.49  &  6.08 \\
run 5 & 0.48  &  4.41  \\
run 6 & 0.40  &  3.78 \\
run 7 & 0.47  &  5.14 \\
\hline
Average & 0.43   & 4.51 \\
\hline
\end{tabular}
\end{center}
\end{table}

\subsection{Bubbles departure diameter measurement}
\label{sec.BDD}
\begin{figure}
\begin{center}
\begin{minipage}{0.19\linewidth}
\center{\includegraphics[width=1\linewidth]{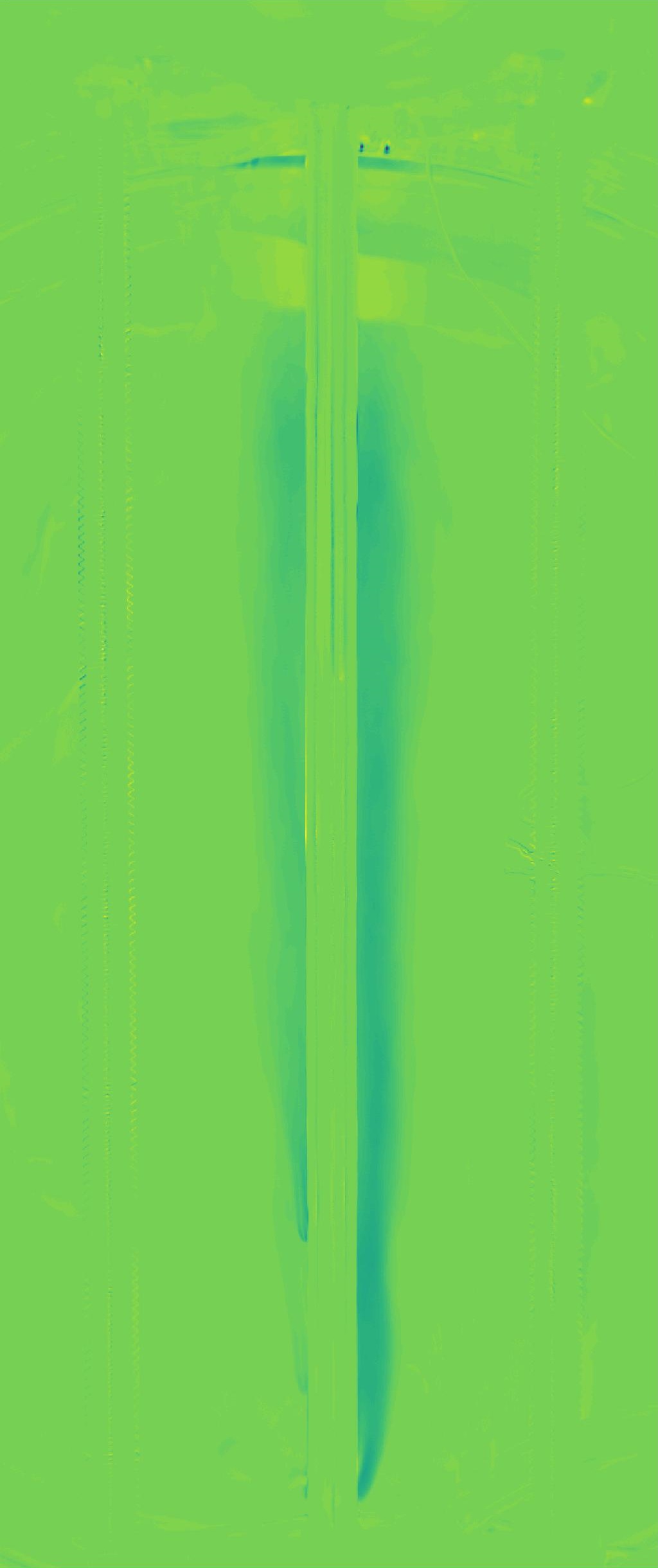}} a)\\
\end{minipage}
\begin{minipage}{0.15\linewidth}
\center{\includegraphics[width=1\linewidth]{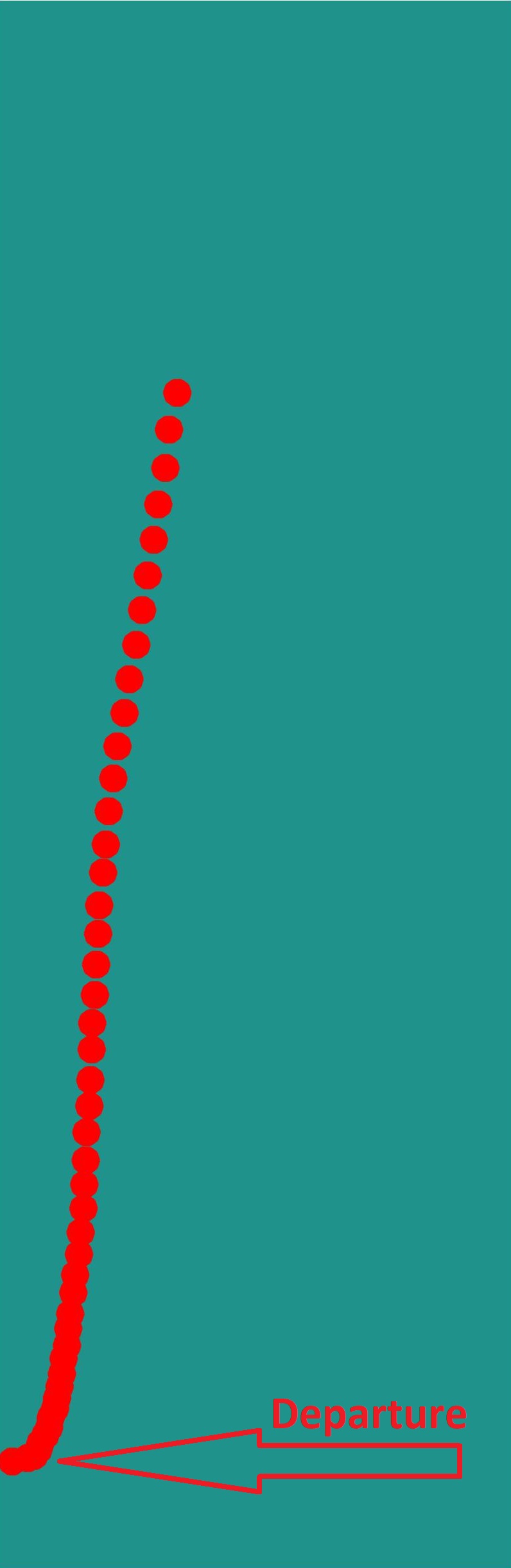}} b)\\
\end{minipage}
\caption{Nucleation sites identification a) and bubble departure definition by center of mass position b)}
\label{fig:NC identification}
\end{center}
\end{figure}

In order to support the obtained findings on HTC, an algorithm was developed to help to extract the data on bubbles departure diameter from high speed recordings. The nucleation sites were first identified by a simple average of all the images recorded during a given experiment. The position and the number of identified nucleation sites are then validated manually. For example, in the Fig. \ref{fig:NC identification}(a), the nucleation sites and the corresponding plumes appear in dark green for the shortest spacing tested.
The software for data processing was developed with GNU Octave and the package “image” \cite{GNUOctave}.
The data processing consists of the following steps:
\begin{enumerate}
\item	A region of interest was defined around each nucleation site.
\item	An image subtraction was carried out for each frame in order to eliminate the background.
\item The image was then readjusted to cover 256 grey levels.
\item Thresholding was applied to limit the impact of measurement noise and to binarize the image.
\item When necessary, the smallest objects (parasitic objects) were eliminated by “bwareaopen” function \cite{GNUOctave}.
\item The “bwlabel” function \cite{GNUOctave} was used to associate a label for connected components of each image.
\item The “regionprops” function, common to many programming languages including GNU Octave \cite{GNUOctave}, was finally used to obtain the data corresponding to each bubble (area, length of main axes, diameter).
 \end{enumerate}
The described algorithm was applied only for two lateral sites to the right and to the left of each tubes i.e. the bubbles interfering with the tubes and with the others bubbles where not considered. Moreover, the bubble departure diameter was defined at the moment when the bubble starts to slide along the tube. It can be clearly seen at the bubble's center of mass position shown in Fig. \ref{fig:NC identification}(b).

\subsection{Uncertainty analysis}

The uncertainty analysis was carried out according to guidelines for evaluating and expressing the uncertainty \cite{taylor1994guidelines}. The maximum expanded standard
uncertainty of the experimental results were evaluated as 0.2 K for wall superheat, 15 W/m$^2$ for heat flux density, 140 W/m$^2\cdot$K for HTC and 0.2 mm for bubble departure diameter.

\section{Results}

\subsection{Long spacings}

\begin{figure}
\centering
\includegraphics[width=0.45\linewidth]{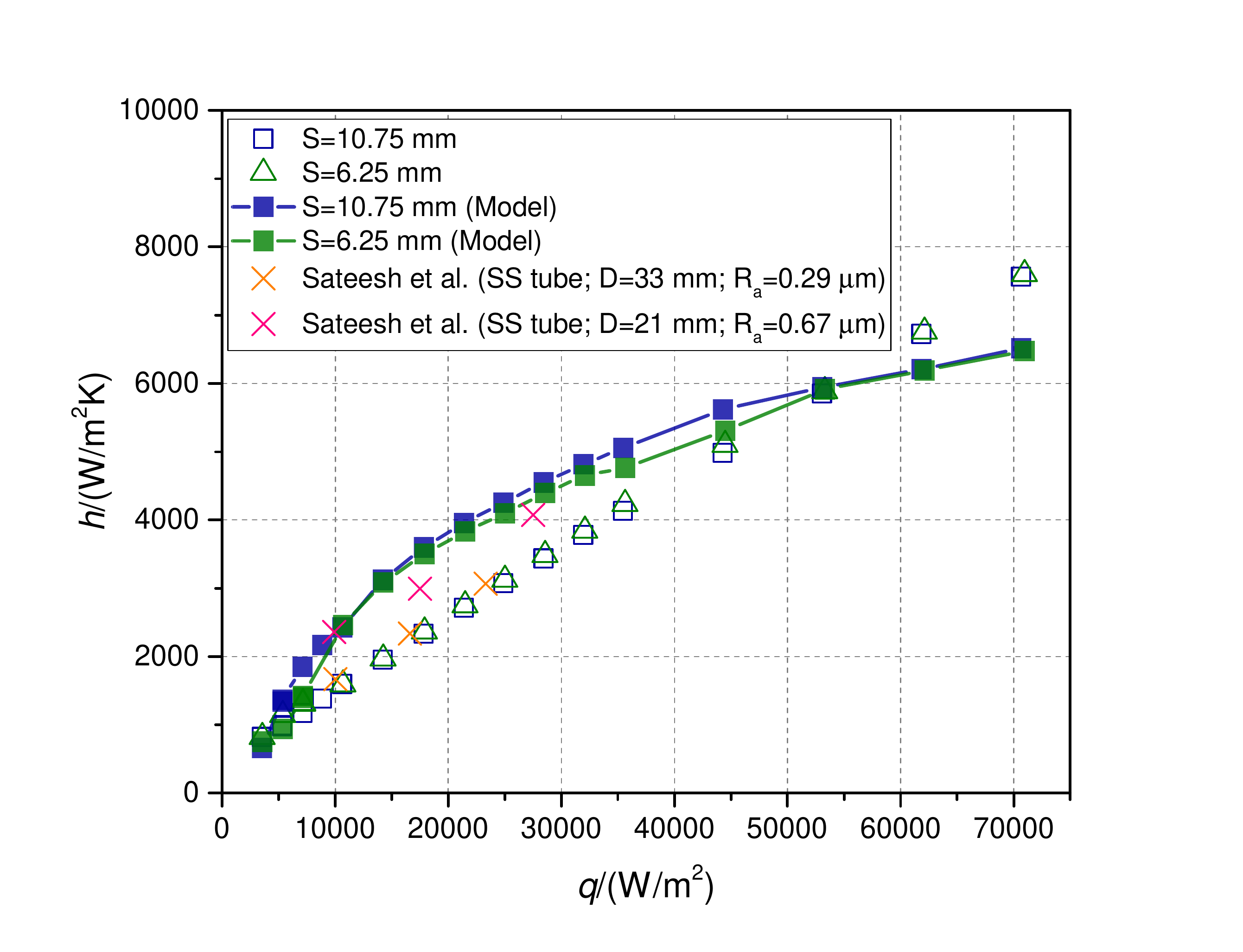}
\caption{Comparison of mean arithmetic HTC at long spacings versus experimental data reported by Sateesh et al. \cite{sateesh2009experimental} and correlation proposed by Jakob and Hawkins \cite{jakob1957elements}}
\label{fig:HTC exp mod mean}
\end{figure}

\begin{figure}
\begin{minipage}[h]{0.45\linewidth}
\center{\includegraphics[width=1\linewidth]{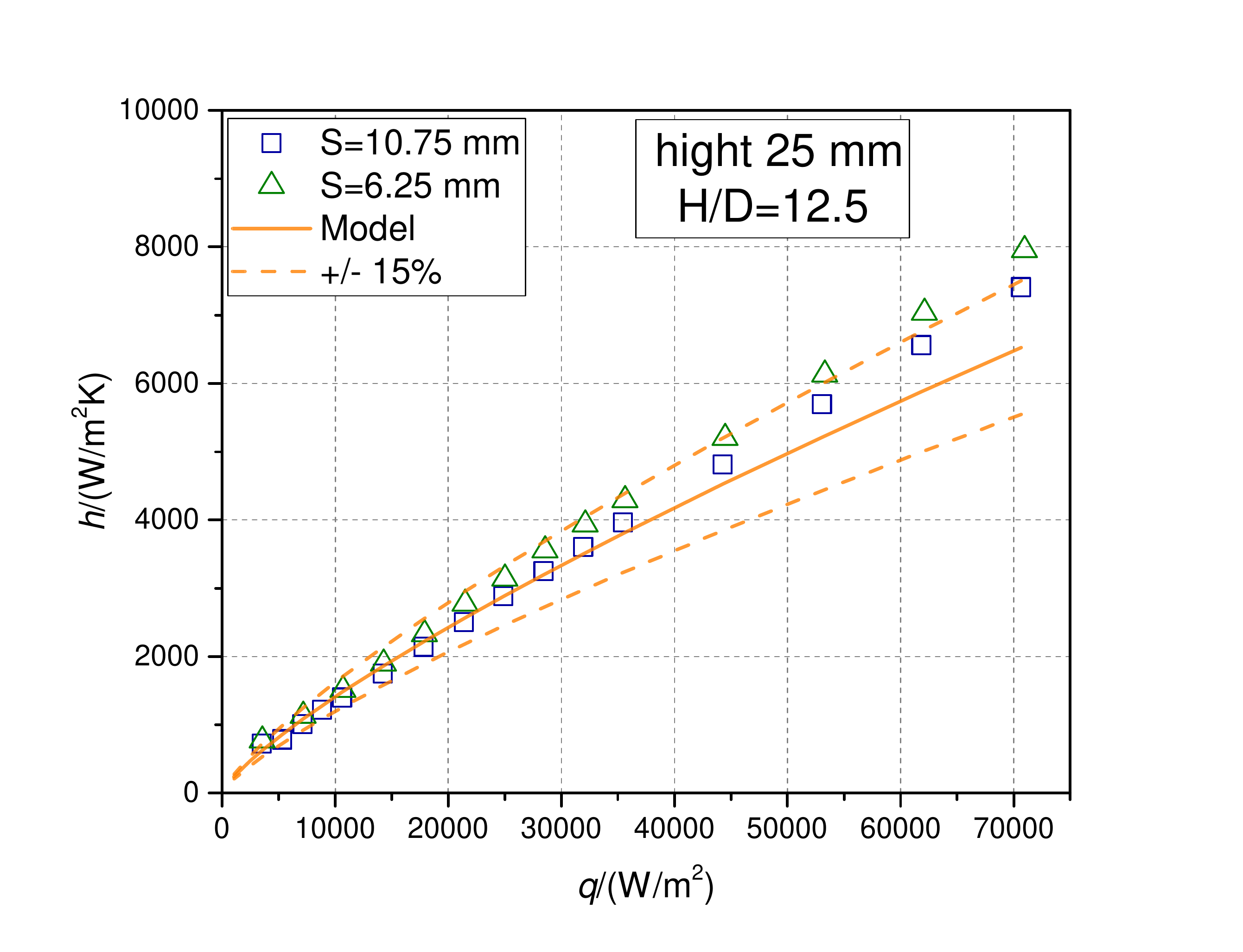}} \\
\end{minipage}
\begin{minipage}[h]{0.45\linewidth}
\center{\includegraphics[width=1\linewidth]{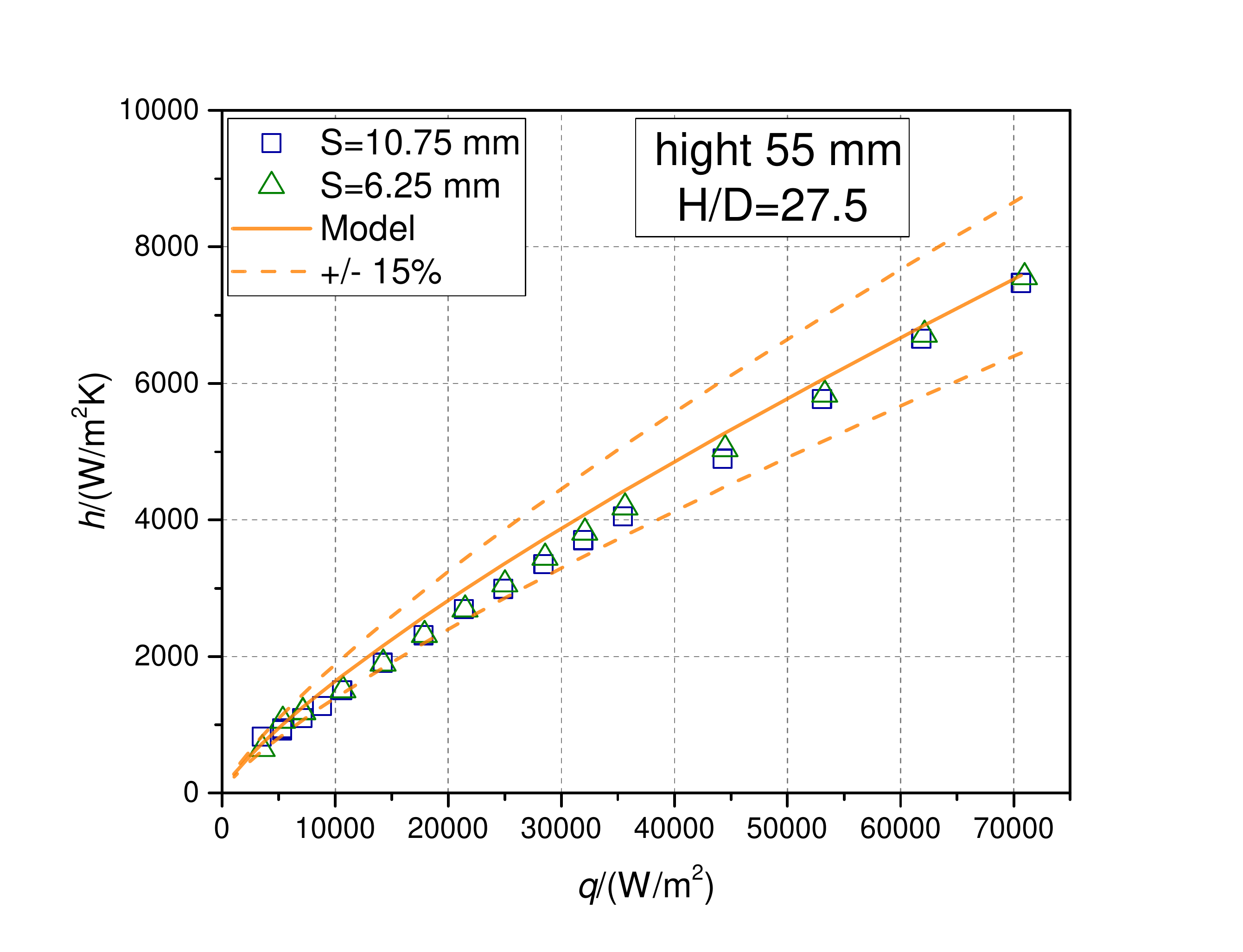}} \\
\end{minipage}
\begin{minipage}[h]{0.45\linewidth}
\center{\includegraphics[width=1\linewidth]{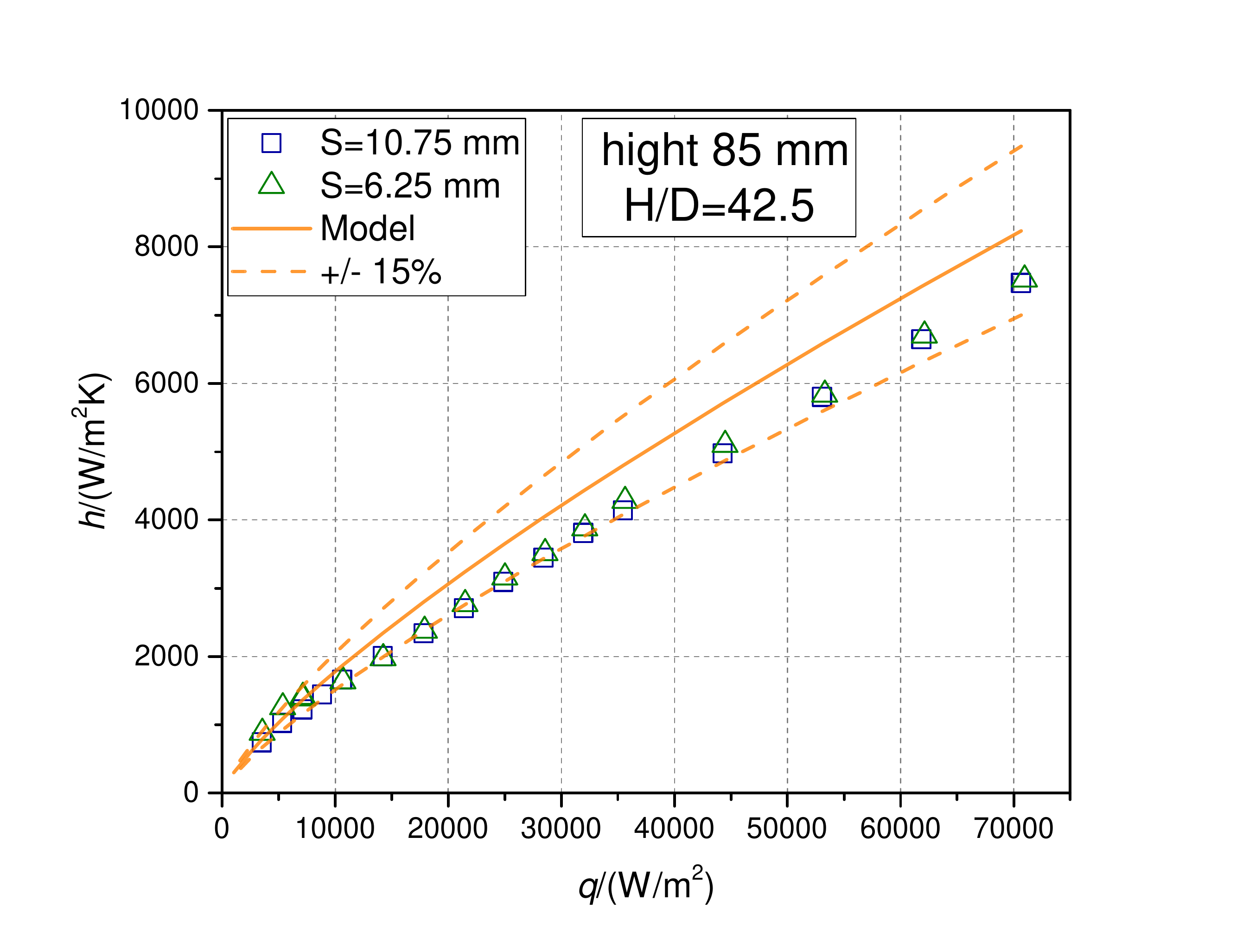}} \\
\end{minipage}
\begin{minipage}[h]{0.45\linewidth}
\center{\includegraphics[width=1\linewidth]{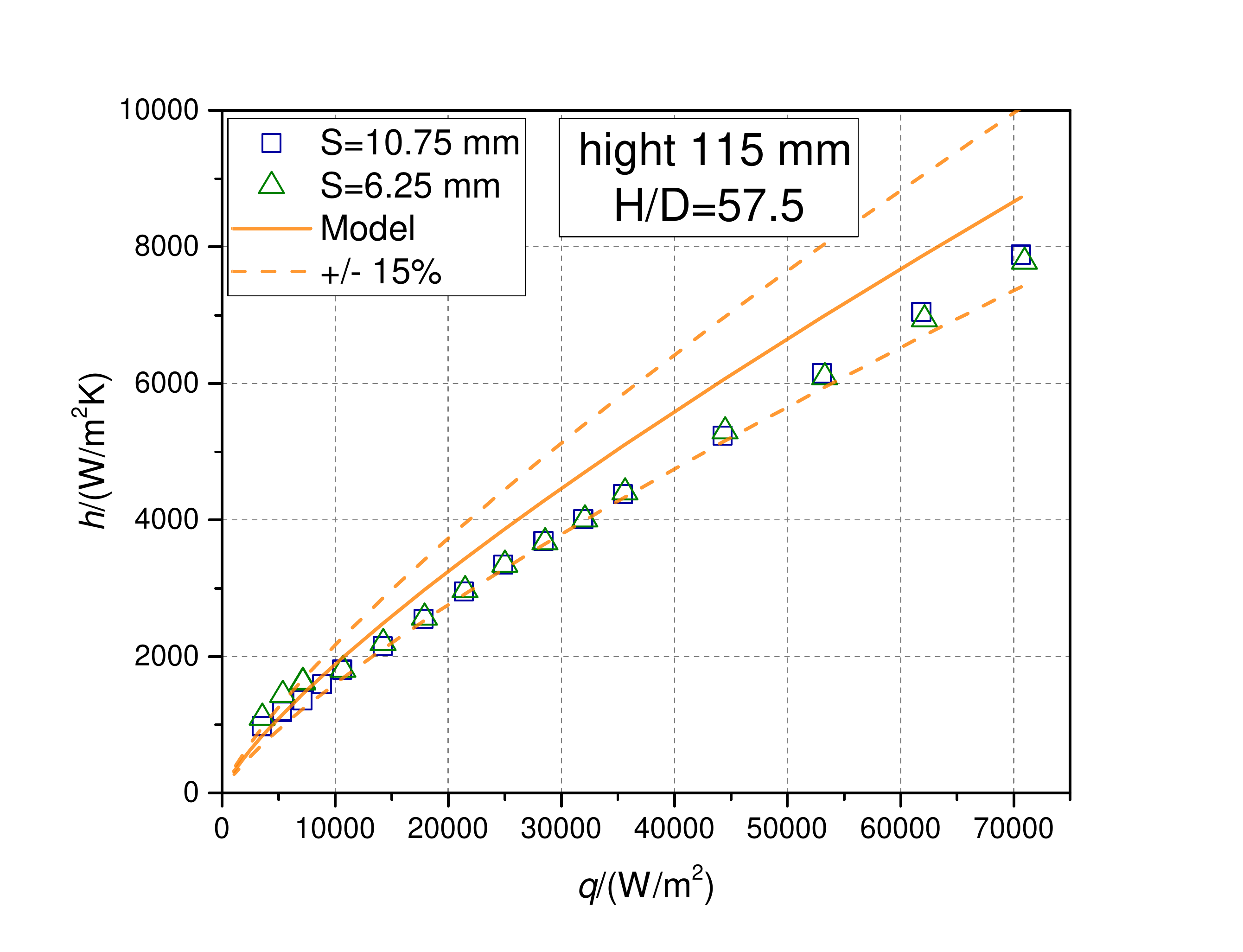}}\\
\end{minipage}
\caption{Comparison of local HTC at long spacings and correlation reported by Tian et al. \cite{tian2018experimental}}
\label{fig:HTC exp mod}
\end{figure}

\begin{figure}
\begin{minipage}[h]{0.18\linewidth}
\center{\includegraphics[width=1\linewidth]{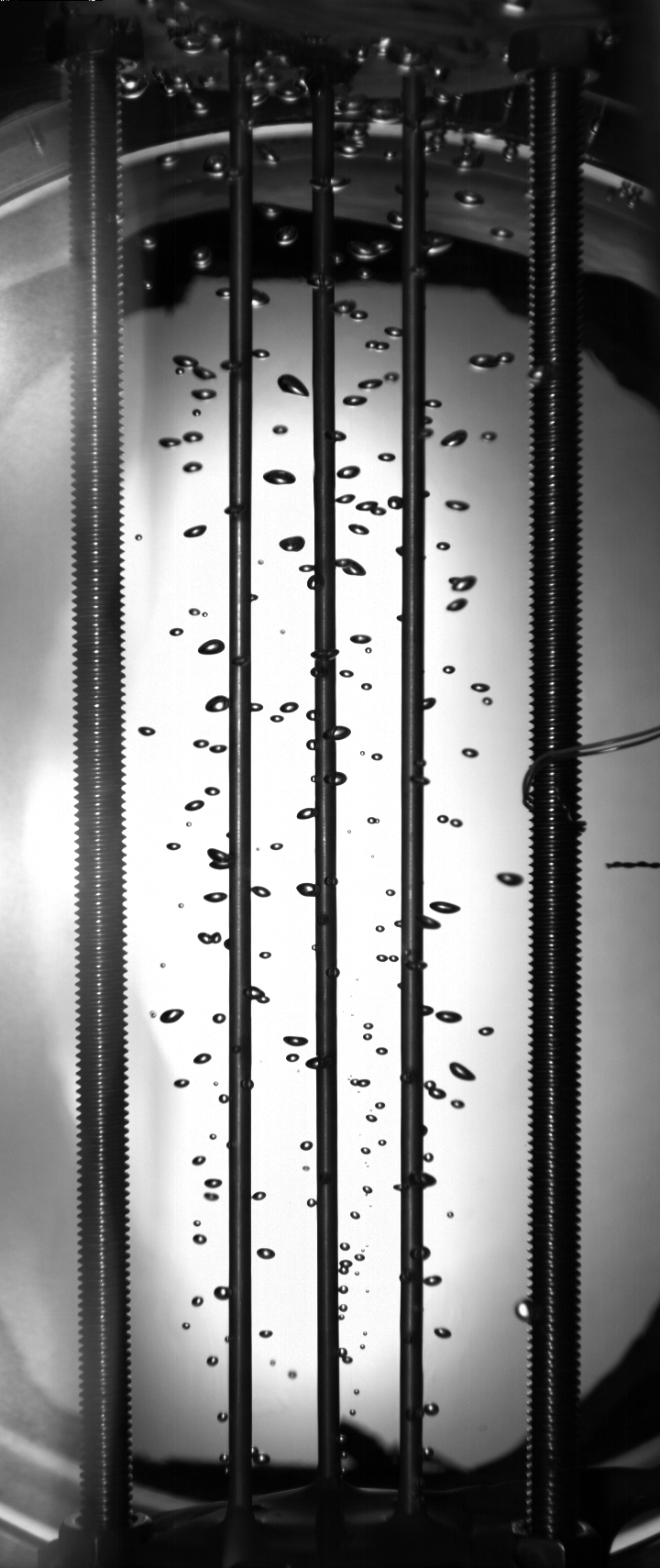}} 7.1 kW/m$^2$ \\
\end{minipage}
\begin{minipage}[h]{0.18\linewidth}
\center{\includegraphics[width=1\linewidth]{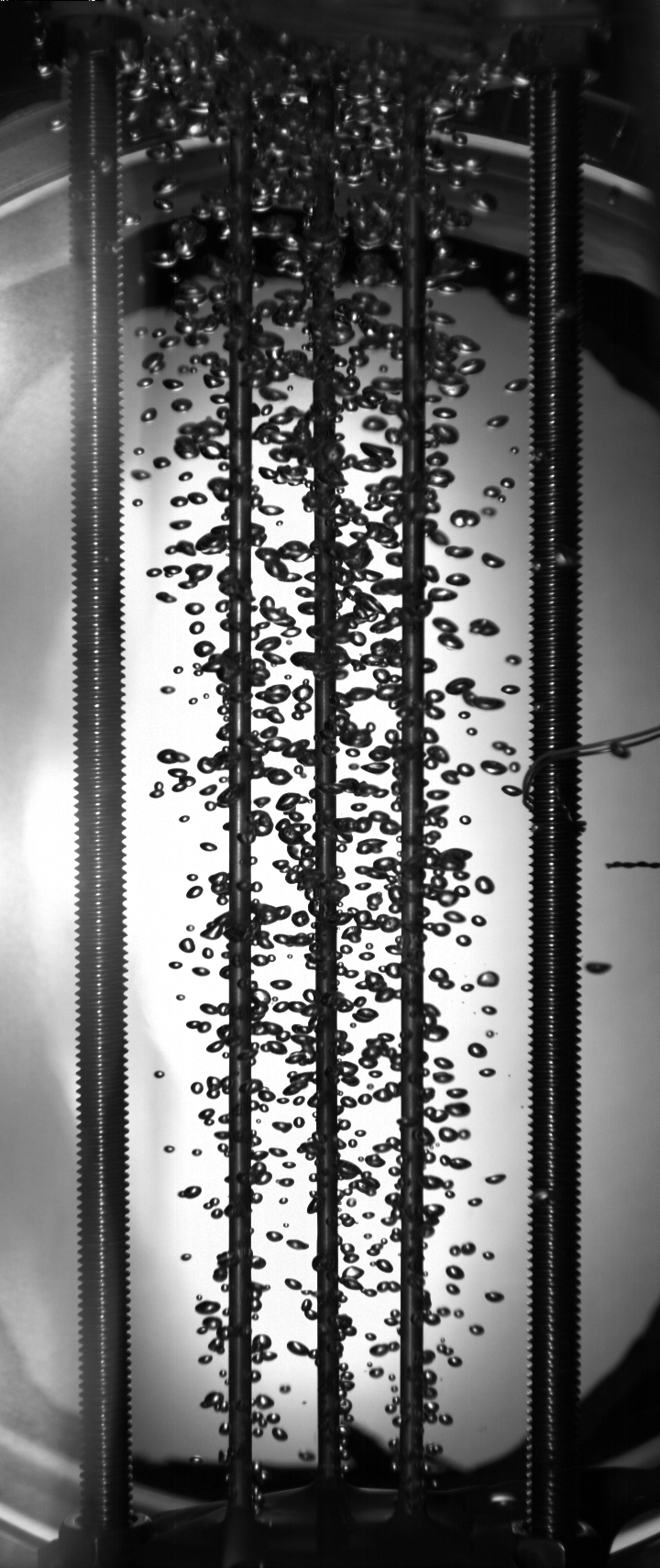}} 17.8 kW/m$^2$ \\
\end{minipage}
\begin{minipage}[h]{0.18\linewidth}
\center{\includegraphics[width=1\linewidth]{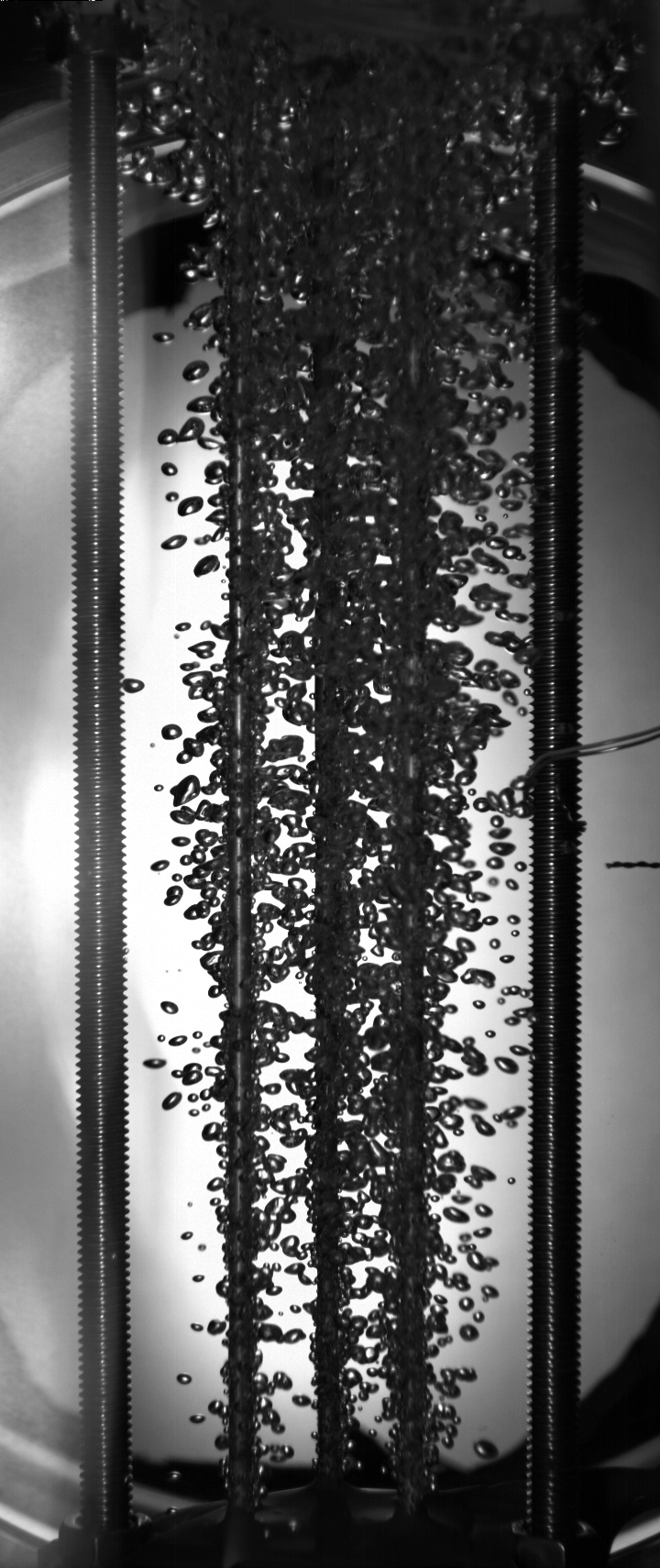}} 35.5 kW/m$^2$ \\
\end{minipage}
\begin{minipage}[h]{0.18\linewidth}
\center{\includegraphics[width=1\linewidth]{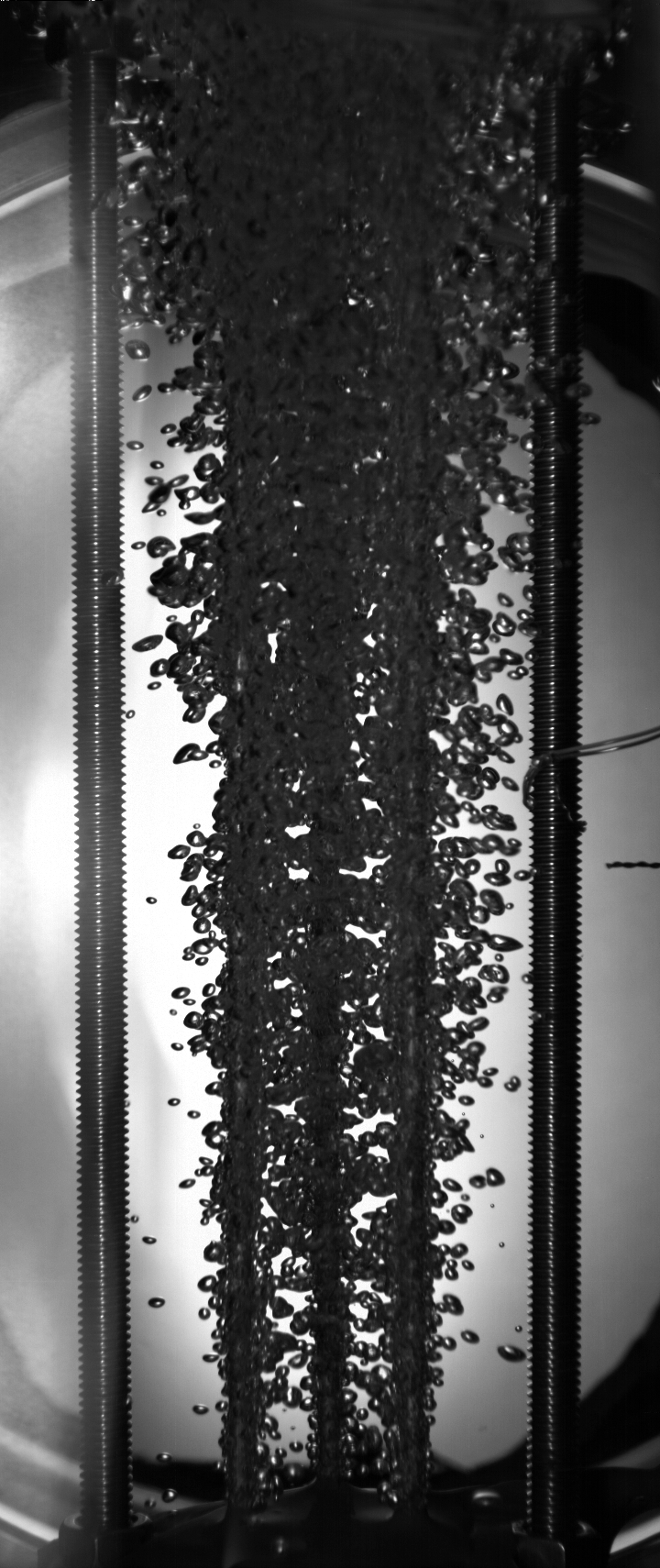}} 53.1 kW/m$^2$ \\
\end{minipage}
\begin{minipage}[h]{0.18\linewidth}
\center{\includegraphics[width=1\linewidth]{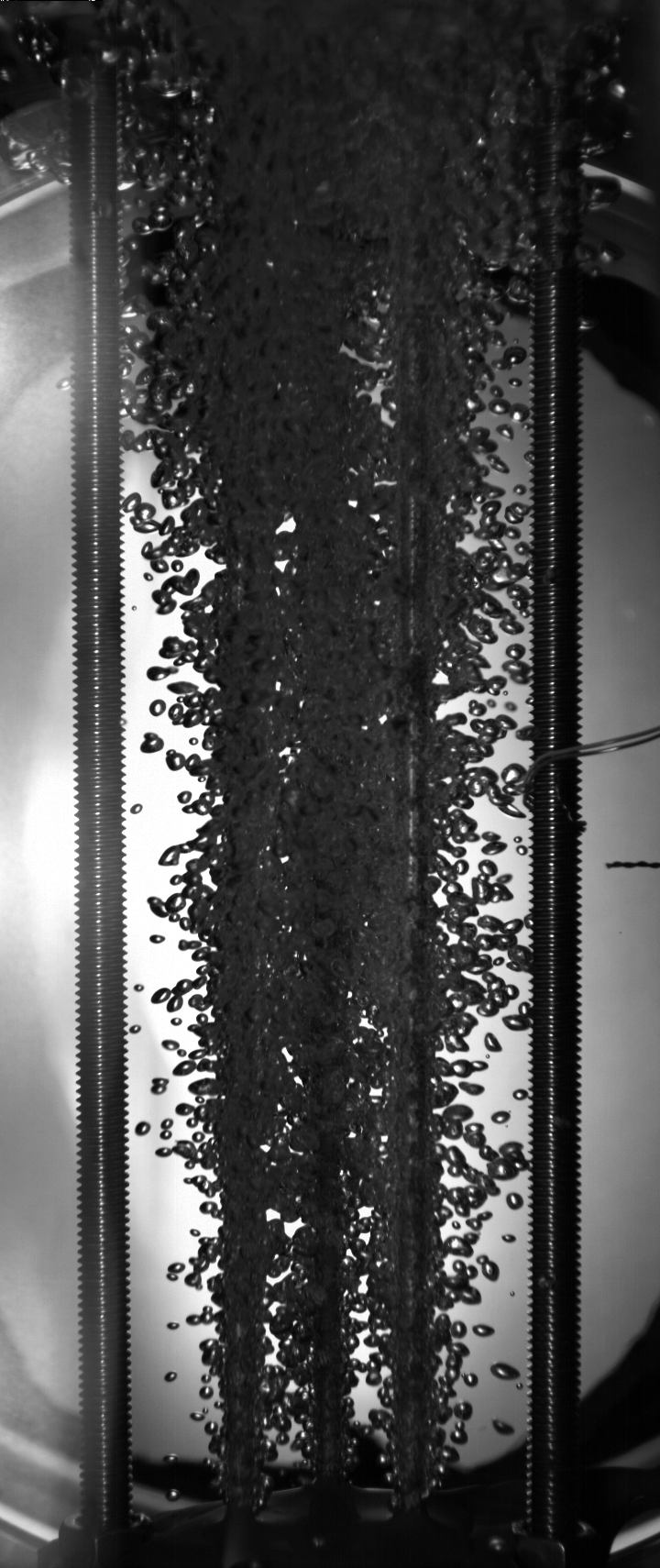}} 70.6 kW/m$^2$ \\
\end{minipage}
\vfill
\begin{minipage}[h]{0.18\linewidth}
\center{\includegraphics[width=1\linewidth]{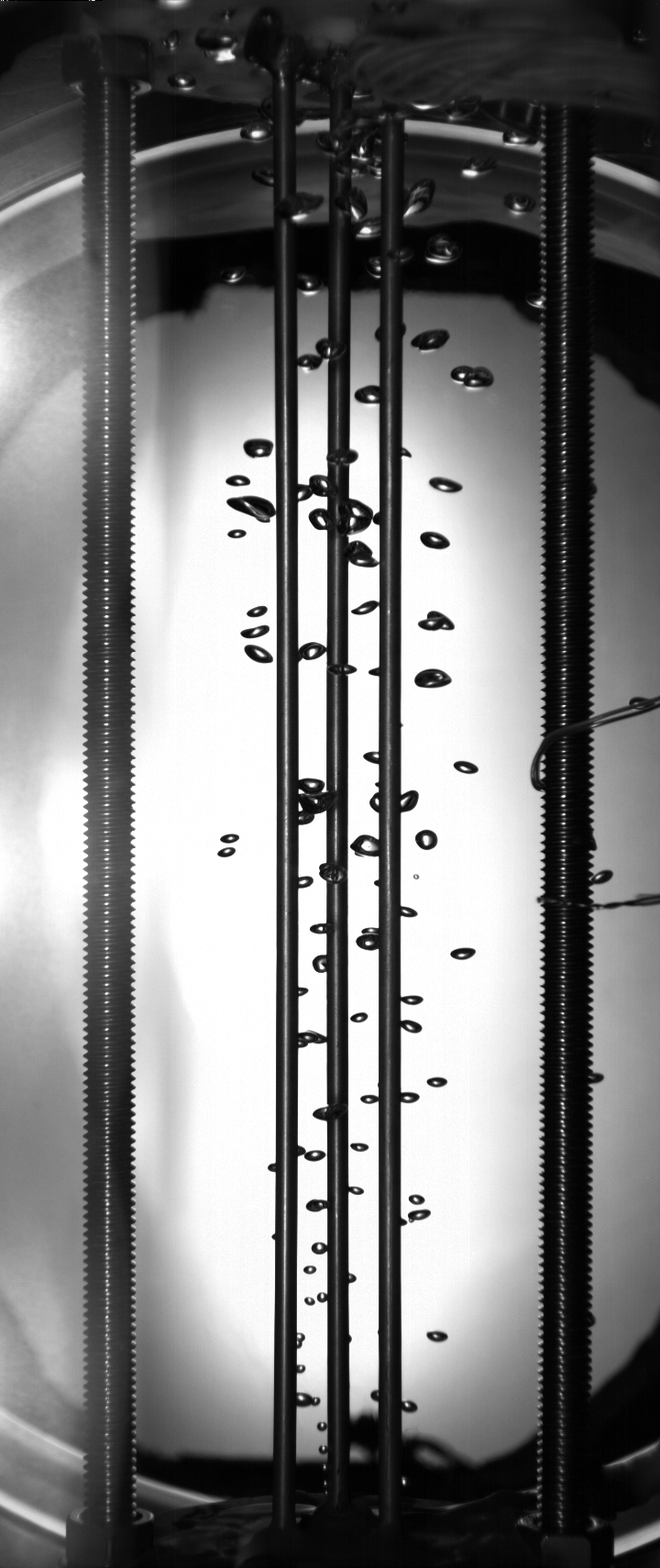}} 7.2 kW/m$^2$ \\
\end{minipage}
\begin{minipage}[h]{0.18\linewidth}
\center{\includegraphics[width=1\linewidth]{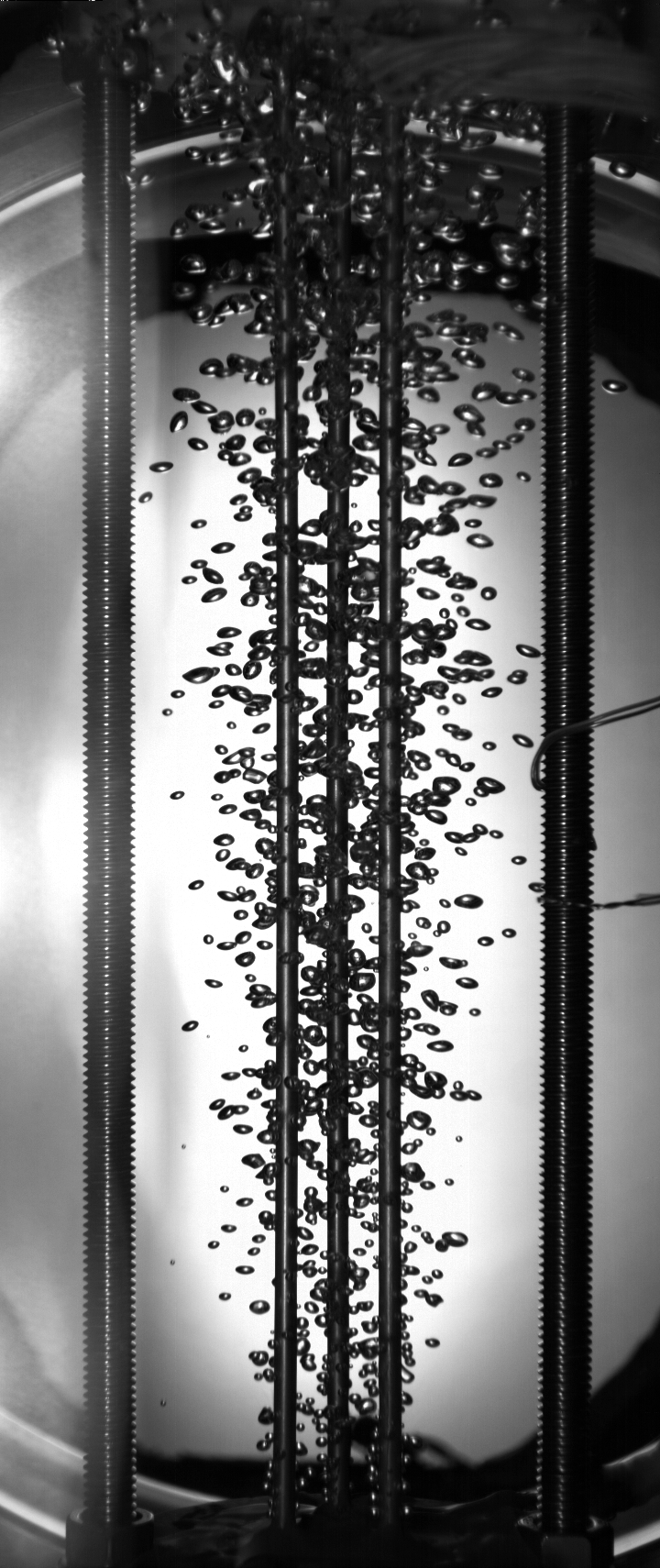}} 17.9 kW/m$^2$ \\
\end{minipage}
\begin{minipage}[h]{0.18\linewidth}
\center{\includegraphics[width=1\linewidth]{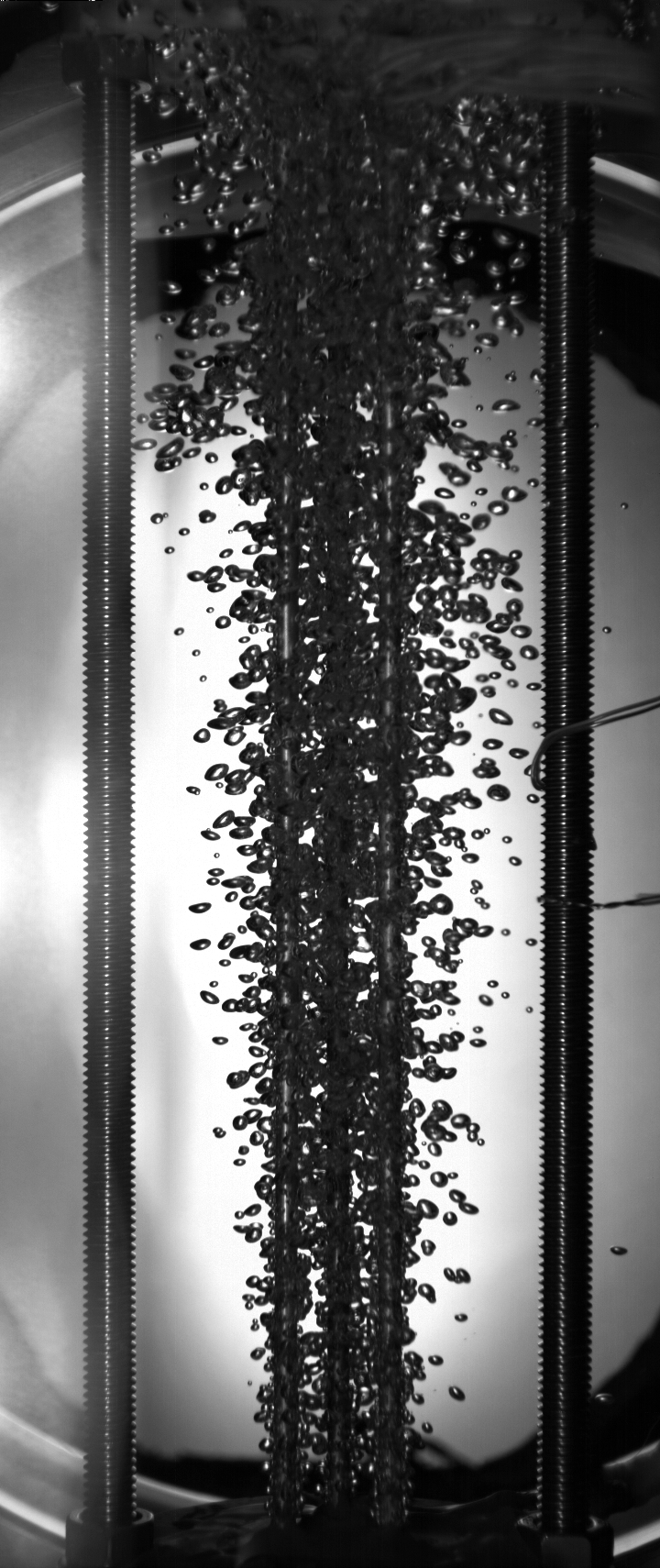}} 35.6 kW/m$^2$ \\
\end{minipage}
\begin{minipage}[h]{0.18\linewidth}
\center{\includegraphics[width=1\linewidth]{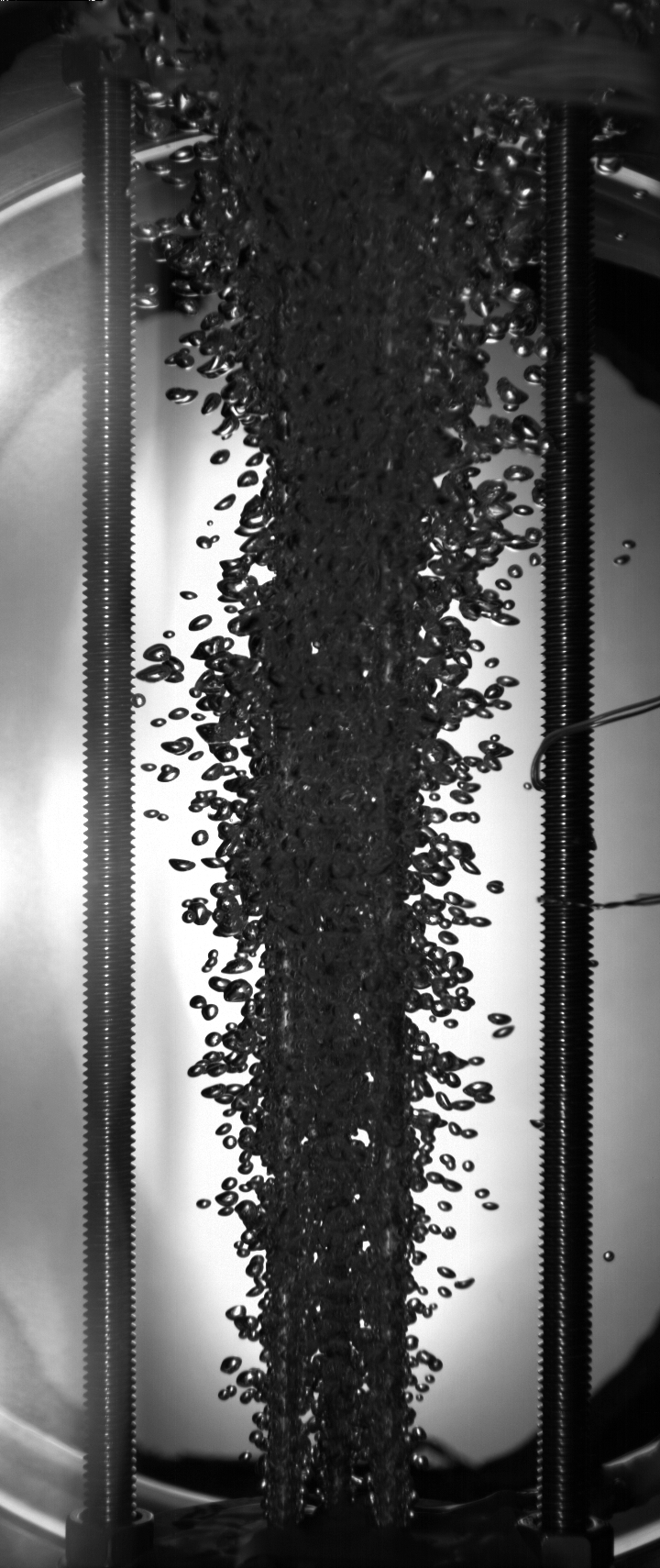}} 53.3 kW/m$^2$ \\
\end{minipage}
\begin{minipage}[h]{0.18\linewidth}
\center{\includegraphics[width=1\linewidth]{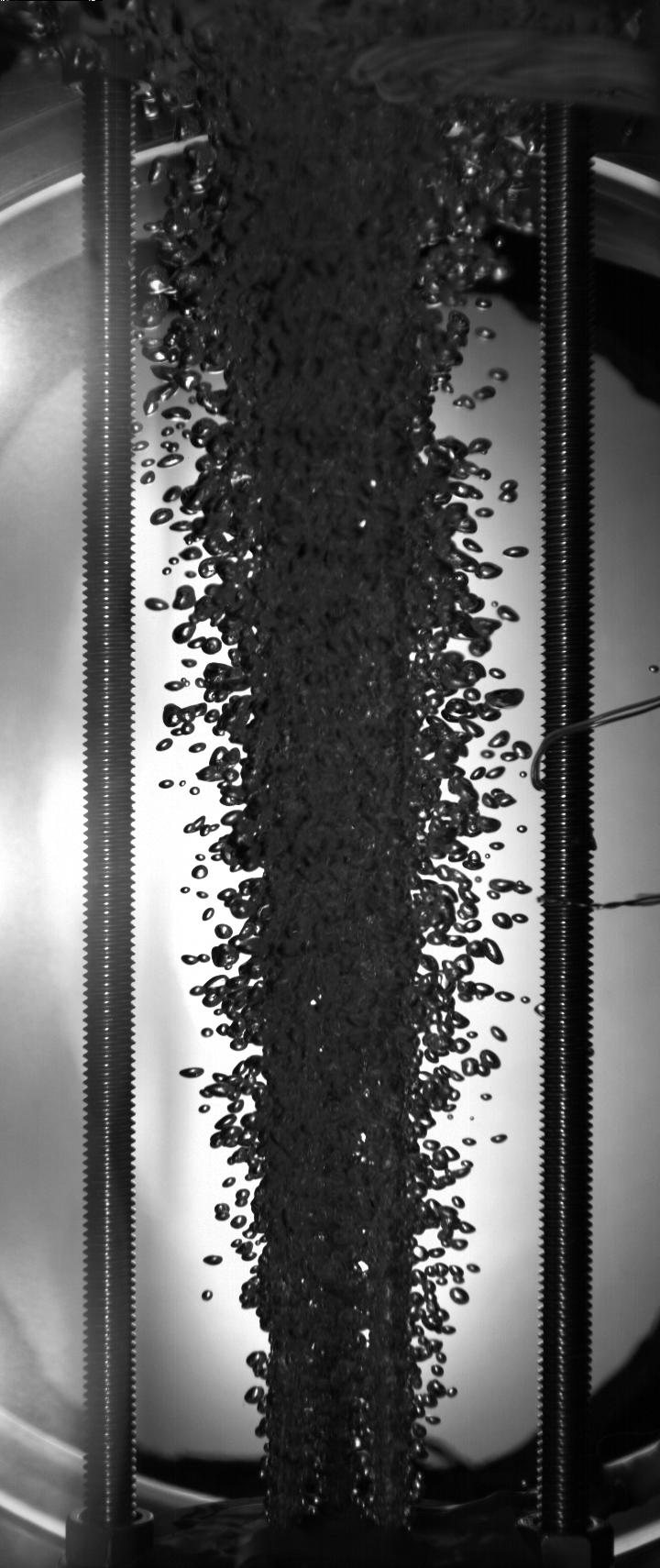}} 71.0 kW/m$^2$ \\
\end{minipage}
\caption{Snapshots of boiling process at different heat fluxes and long spacings: Top row $S=10.75$; Bottom row $S=6.25$.}
\label{fig:SP3 and SP2}
\end{figure}

The experiments were conducted at a descending order of spacing given in Table \ref{tabular:spacing}. The long spacings of 10.75 and 6.25 mm show almost identical results in a sense of HTC (see Fig. \ref{fig:HTC exp mod mean} and \ref{fig:HTC exp mod}) and bubble dynamics (see Fig. \ref{fig:SP3 and SP2}). The nucleation of bubbles occurs preferably at the bottom part of the bundle, that was also noticed in several studies \cite{chun1998effects,gupta2010nucleate,chung2015heat}. This is due to thermal boundary layer destruction by rising up bubbles. It can also be seen, that the bubbles are creating a "flambeau"-like or "cone"-like shape plum along the tubes height that is turned by its apex to the base of the tubes bundle. Similar effect was previously reported and described for the bubbles released in the liquid pool \cite{uchiyama2015numerical,lima2016influence,fraga2016influence}. Additionally, to the reported findings, while boiling, this effect has been also caused by the interaction between floating up bubbles and the bubbles on the stage of nucleation and departure. During the nucleation and departure from the vertical surface, bubbles have an impulse in a horizontal plane that can be consequently transmitted to the nearest bubbles. Hence, the observed "flambeau" effect is stronger at higher heat fluxes and shorter spacings, when the number of active nucleation sites is high and they are close to each other. Nevertheless, the "flambeau" effect tends to reduce the interaction between the vertical surface and bubbles and its impact on heat transfer is insignificant. In other words, the bundles with long spacings behave similarly to the single tube that also confirm findings of Kang \cite{kang2016effect}. 

Unfortunately, there is no suitable correlation to predict the HTC during pool boiling of the case studied here in terms of boiling fluid and surface properties, surface orientation and roughness etc. \cite{gorenflo2014prediction,stephanvdi,stephan1992heat,thome2006prediction,pioro1999experimental}. Nonetheless, with the comparative purpose, the local HTC values at $S=10.75$ and $S=6.25$ were compared with the correlation $h=0.6425\cdot q^{0.7832}(H/D)^{0.1904}$ proposed by Tian et al. \cite{tian2018experimental} and the mean arithmetic HTC values were compared with experimental data obtained by Sateesh et al. \cite{sateesh2009experimental} and Jakob and Hawkins correlation $h=7.96 \cdot{\Delta T}^3$  \cite{jakob1957elements}. As can be seen in Fig. \ref{fig:HTC exp mod mean}, both experimental data measured here and by Sateesh et al. \cite{sateesh2009experimental} for the pool boiling of ethanol on vertical tube made of SS with $D=33$ mm and $R_a=0.29$ $\mu$m agree very well (within $\pm$5\%). However, the discrepancies are quit high as compared to the HTC data reported for the pool boiling of ethanol on vertical SS tube with  $D=21$ mm and $R_a=0.67$ $\mu$m \cite{sateesh2009experimental} that go up to 50\% at $q\sim10$ kW/m$^2$ and up to 17\% at $q\sim40$ kW/m$^2$. Such inconsistency is probably caused by the difference in roughness and topology of boiling surfaces. Correlations of Tian et al. \cite{tian2018experimental} and correlation of Jakob and Hawkins \cite{jakob1957elements} were proposed for vertically aligned tubes and originally developed using experimental data for pool boiling of water. Surprisingly, the slope and absolute values of local HTC match well with empirical correlation of Tian et al. \cite{tian2018experimental} (see Fig. \ref{fig:HTC exp mod}). On the contrary, both the slope and absolute values of mean arithmetic HTC substantially do not match the correlation of Jakob and Hawkins \cite{jakob1957elements} with the maximum deviations up to 60\% (see Fig. \ref{fig:HTC exp mod mean}). However, considering the simplicity of both correlations \cite{tian2018experimental,jakob1957elements} it is hard to draw any conclusions from a good forecast of Tian et al. correlation \cite{tian2018experimental} and worth of Jakob and Hawkins correlation \cite{jakob1957elements}.

\begin{figure}
\begin{minipage}[h]{0.45\linewidth}
\center{\includegraphics[width=1\linewidth]{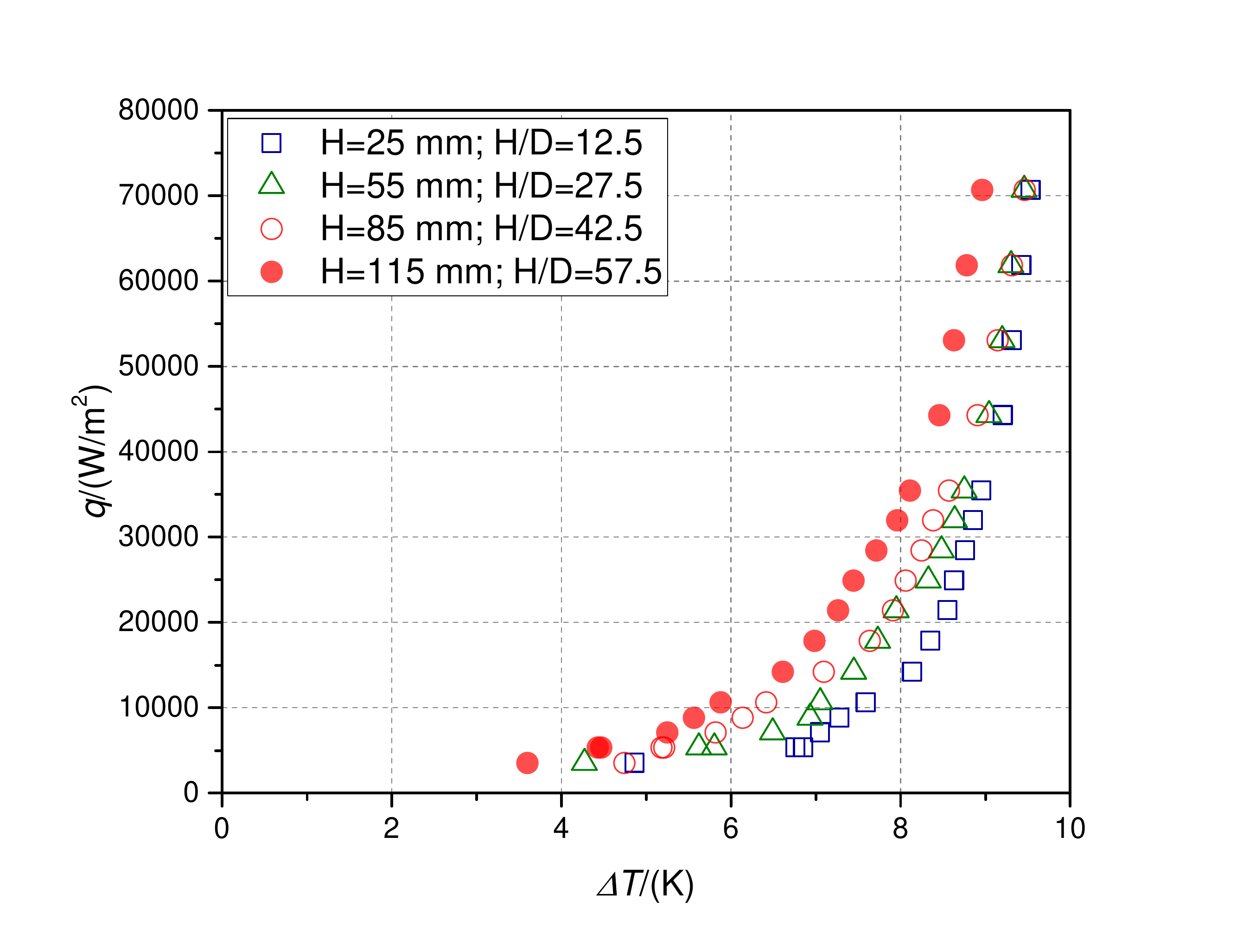}} a) \\
\end{minipage}
\begin{minipage}[h]{0.45\linewidth}
\center{\includegraphics[width=1\linewidth]{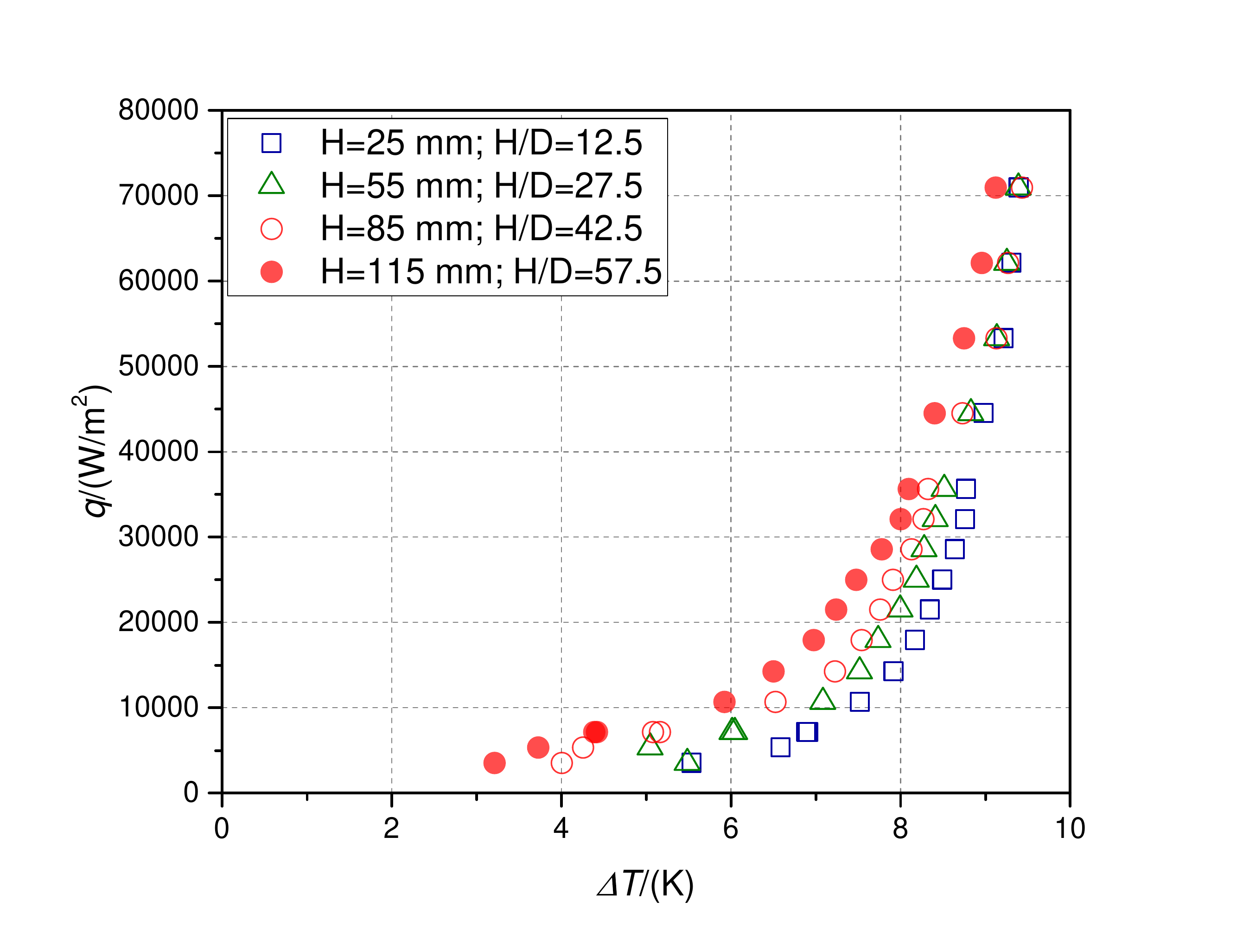}} b) \\
\end{minipage}
\caption{Heat flux density versus wall superheat at different heights for long spacings: a) $S=10.75$ mm, b)  $S=6.25$ mm.}
\label{fig:dT vs H}
\end{figure}

\begin{figure}
\begin{minipage}[h]{0.45\linewidth}
\center{\includegraphics[width=1\linewidth]{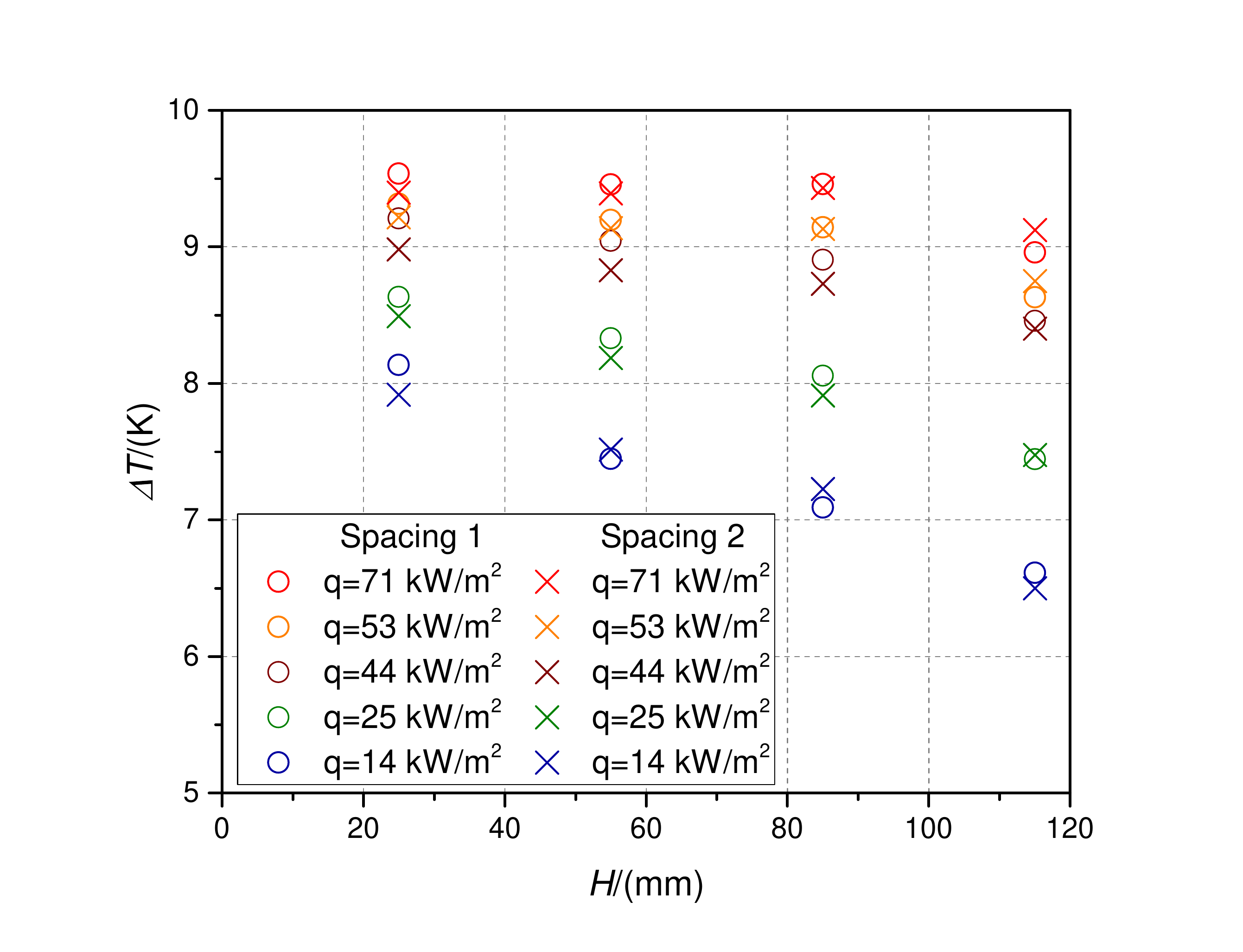}} a) \\
\end{minipage}
\begin{minipage}[h]{0.45\linewidth}
\center{\includegraphics[width=1\linewidth]{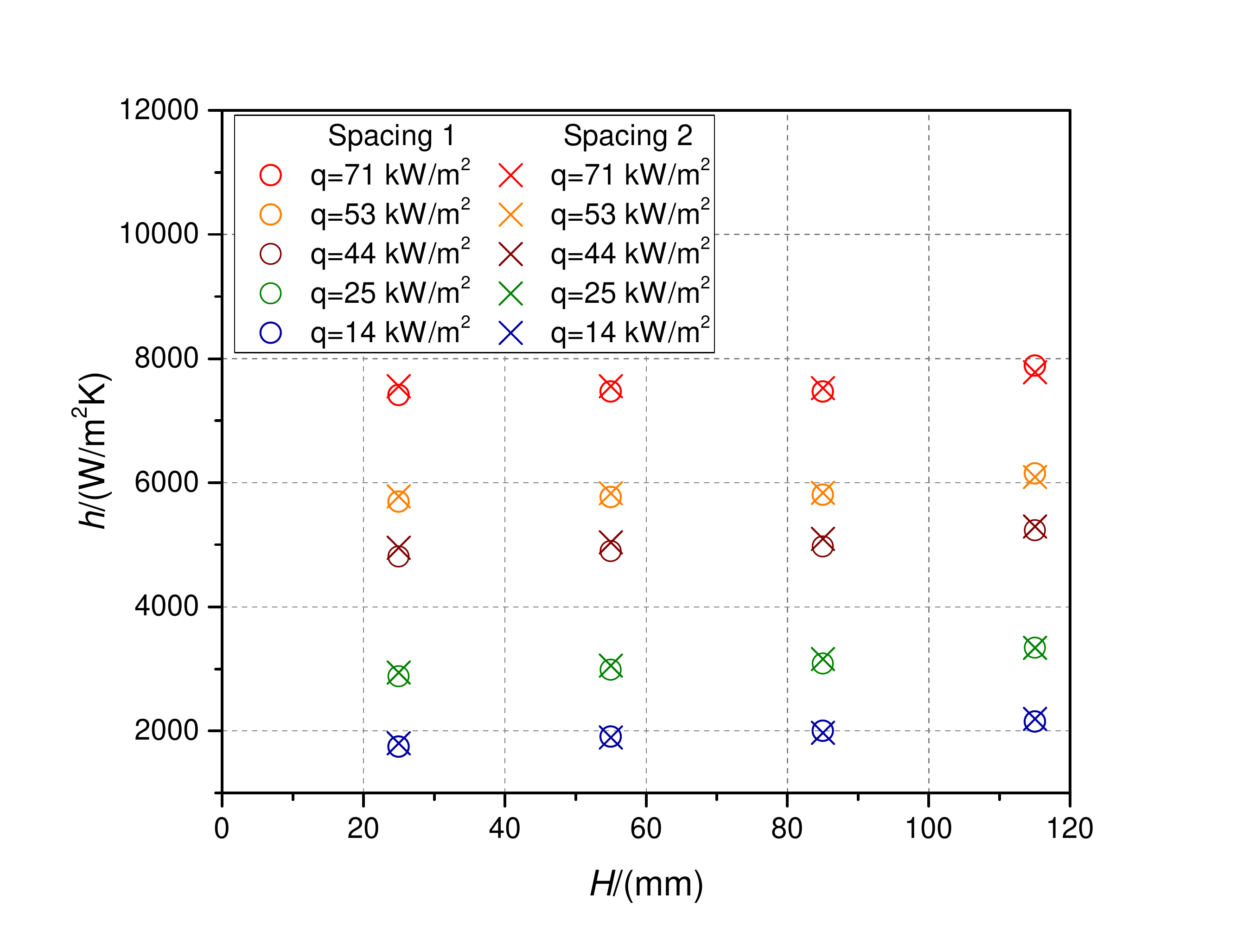}} b) \\
\end{minipage}
\caption{Wall superheat a) and HTC b) versus tubes height.}
\label{fig:dT and h vs H at q}
\end{figure}

The local boiling curves obtained for long spacings are shown in Fig. \ref{fig:dT vs H}. As can be seen, the boiling curve is shifting to the left side versus the height of the tubes. This behavior is typical for boiling on vertical tubes \cite{gupta2010nucleate,chung2015heat, tian2018experimental} that is associated with increased degree of turbulence along side  vertical tube provoked by floating upward bubbles. This effect is pronounced at low heat fluxes, when the convective component of HTC is high (see Fig. \ref{fig:dT and h vs H at q}). However, at high heat fluxes, the main role in the boiling process play bubble nucleation, grow  and departure that equalise the wall superheat and HTC along the tubes height (see Fig. \ref{fig:dT and h vs H at q}).

\subsection{Short spacings}

\begin{figure}
\begin{minipage}[h]{0.45\linewidth}
\center{\includegraphics[width=1\linewidth]{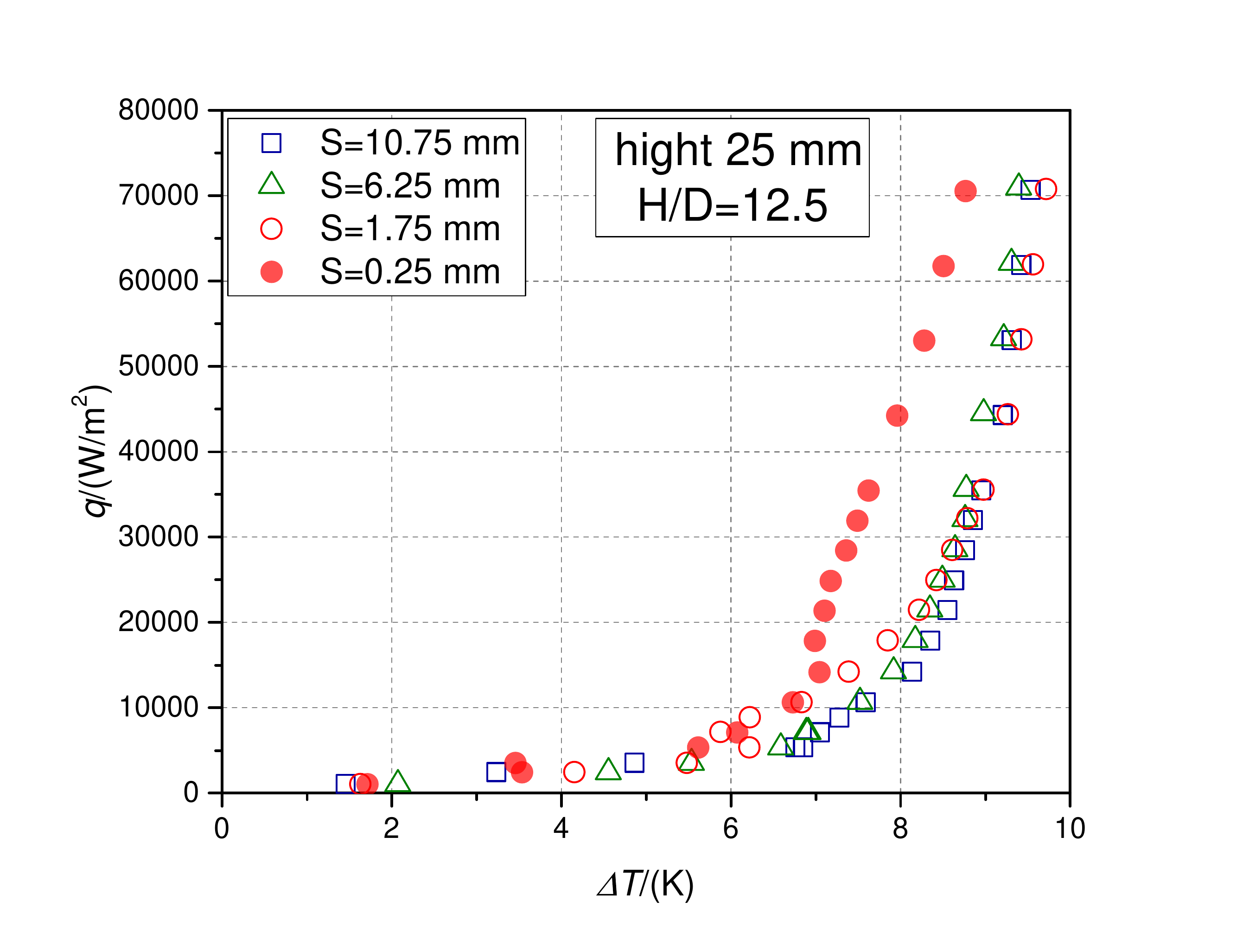}} \\
\end{minipage}
\begin{minipage}[h]{0.45\linewidth}
\center{\includegraphics[width=1\linewidth]{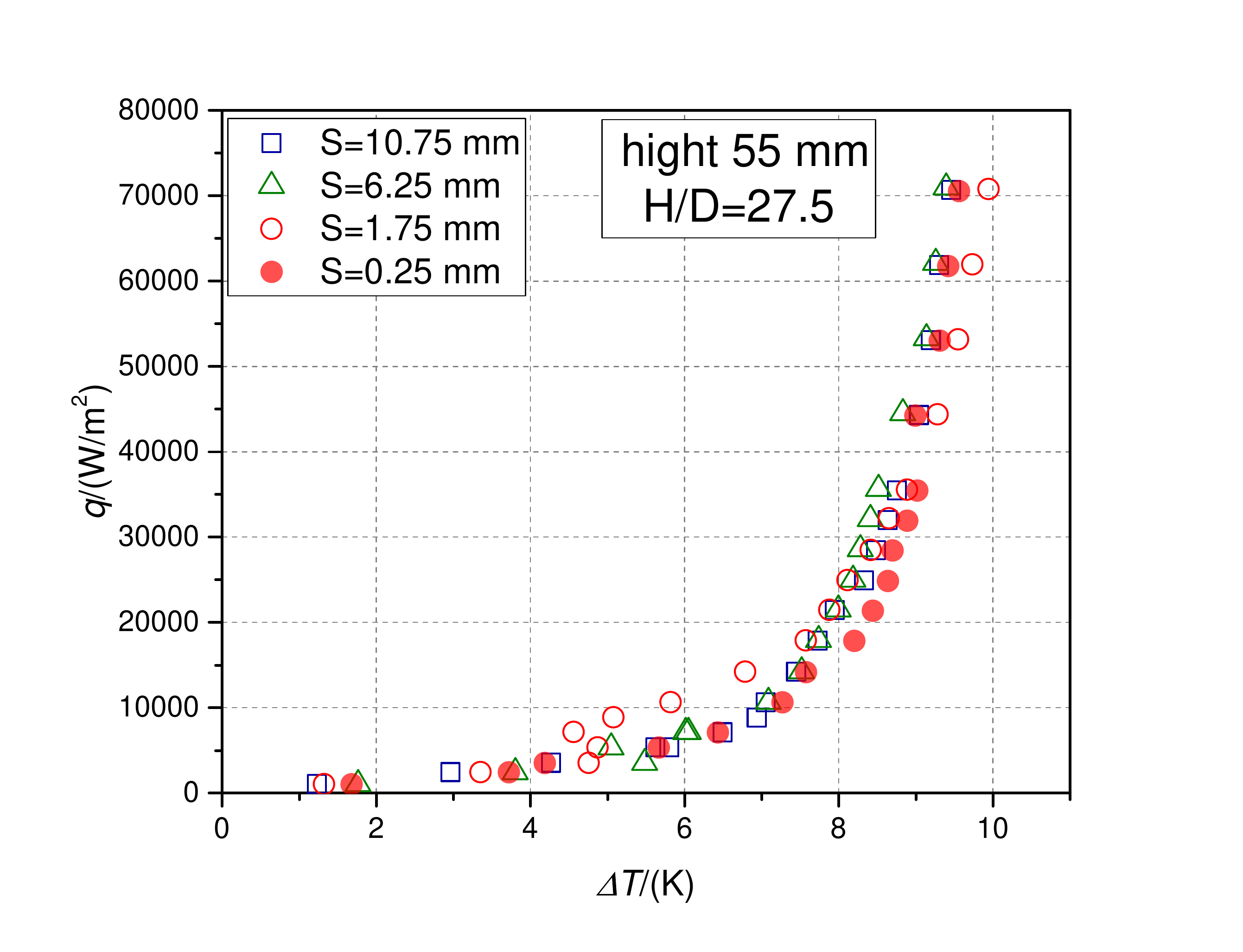}} \\
\end{minipage}
\begin{minipage}[h]{0.45\linewidth}
\center{\includegraphics[width=1\linewidth]{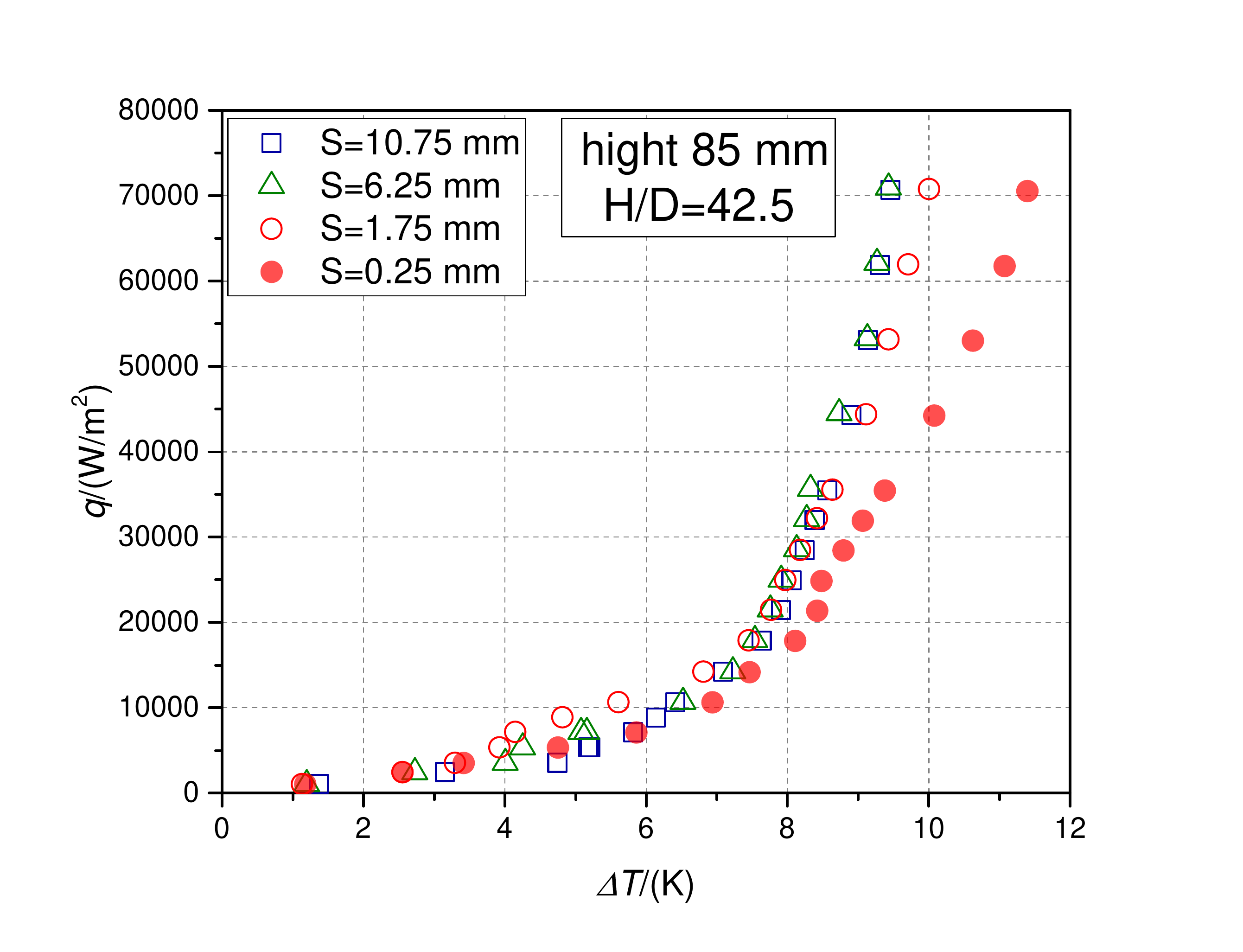}} \\
\end{minipage}
\begin{minipage}[h]{0.45\linewidth}
\center{\includegraphics[width=1\linewidth]{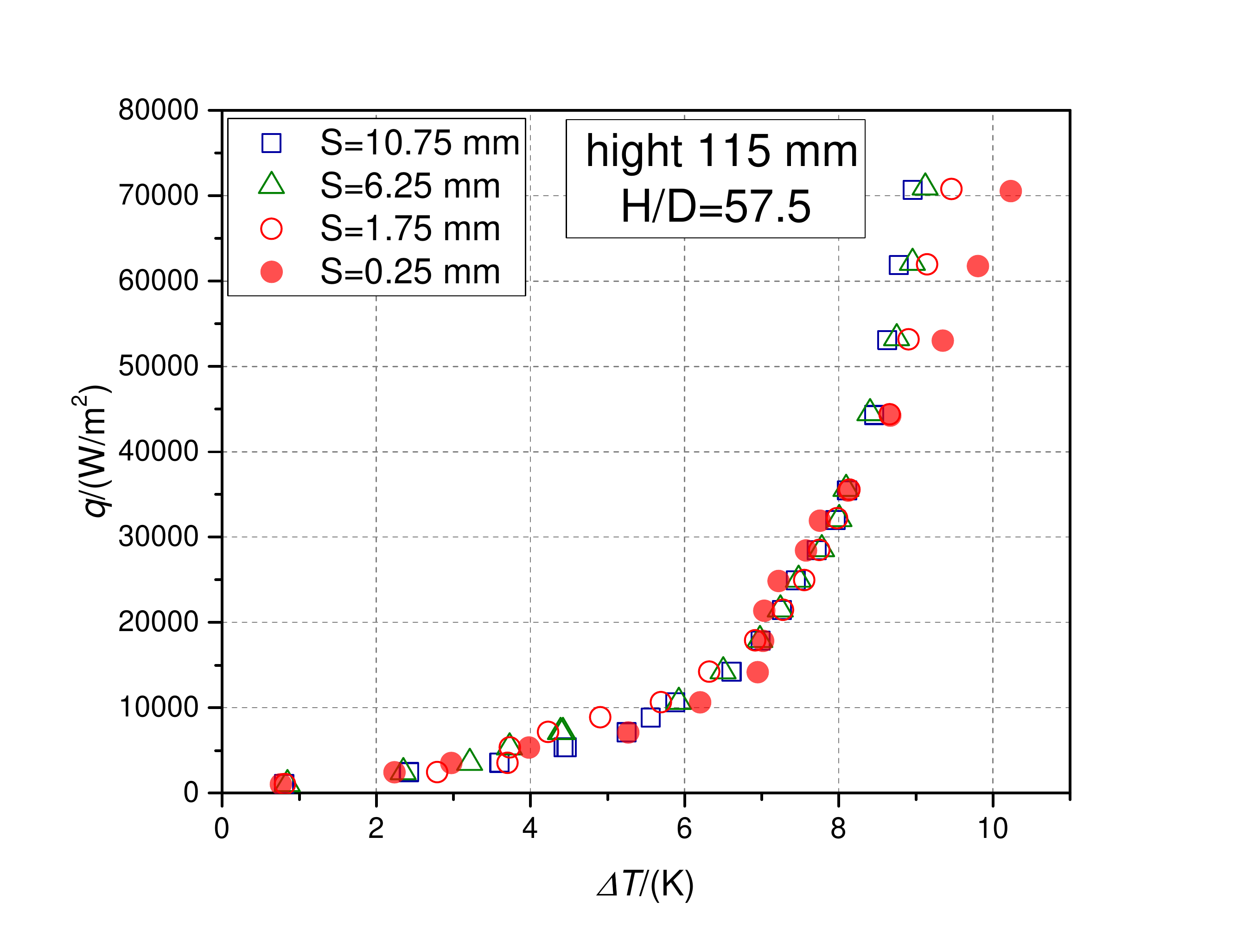}}\\
\end{minipage}
\caption{Heat flux density versus wall superheat and spacings}
\label{fig:dT}
\end{figure}

\begin{figure}
\centering
  \includegraphics[width=0.4\textwidth]{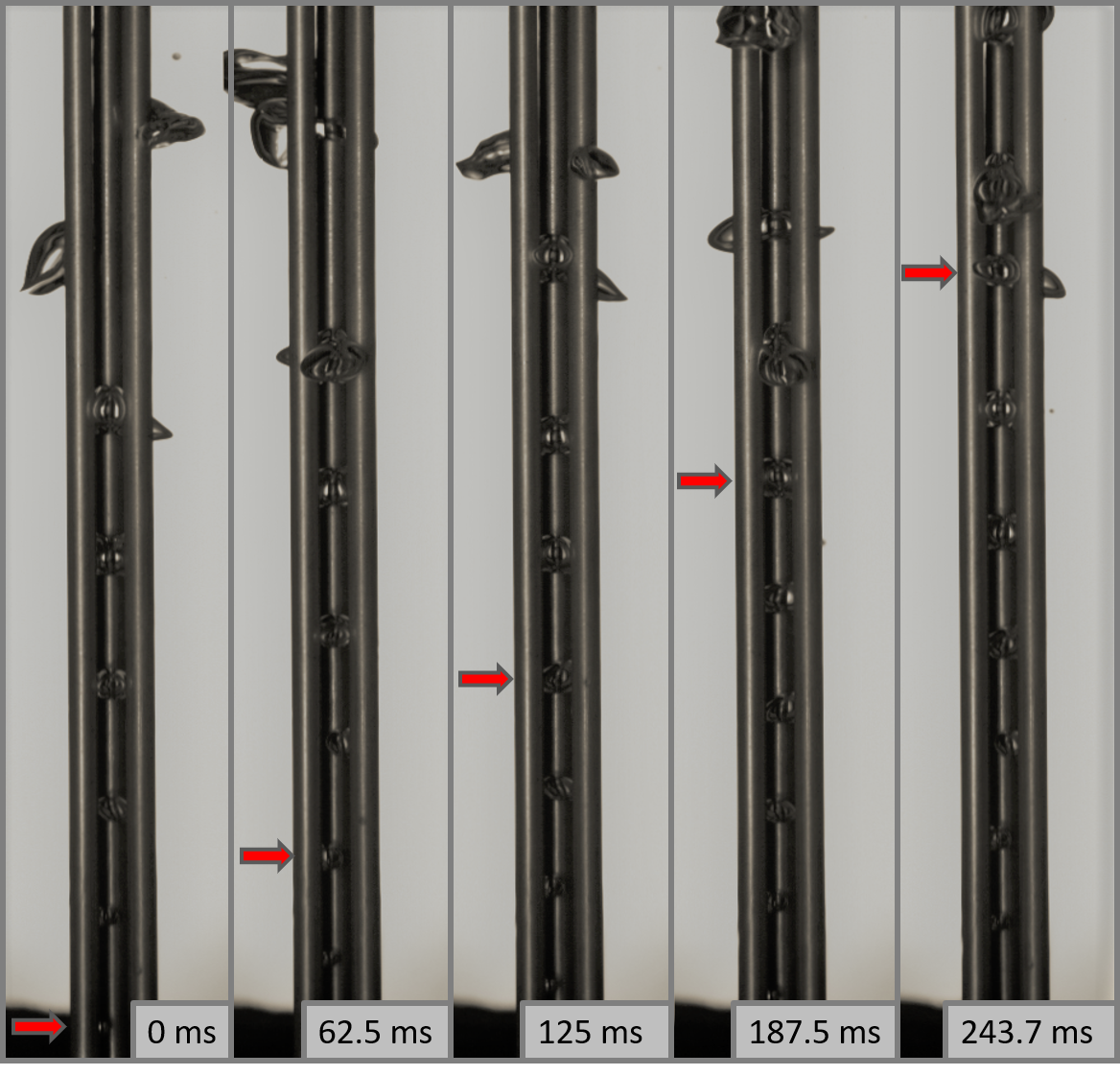}  \\
\caption{Processed snapshots of one slug formation-destruction cycle  at $q=7.1$ kW/m$^2$ and $S=1.75$ (see also movie 1 in supplementary material)}
\label{fig:Slug}
\end{figure}

Further reduction of spacing to $S=1.75$ mm has caused slug flow on the bubbles nucleated in the gap created by three tubes at low heat flux density (up to 20 kW/m$^2$) when the boiling process is not fully developed. This mode of bubbles helps to cool down the tubes efficiently. However, the impact of slugs on heat transfer vary over the height of the tubes. The observations show, after the bubble nucleation, the bubble starts to grow and gradually turn into a slug. During this time the interaction between the bubble and the walls of the tubes increases and as the result, the wall superheat decreases. That explain that the wall superheat at H/D of 12.5 is higher than at H/D of 27.5 (see Fig. \ref{fig:dT}). At a certain height, the slug reaches it's maximum size and finally slides out of a gap and breaks into several smaller bubbles. After that, those smaller bubbles continue to float upward and gradually move apart from the tubes. Thus, their influence on heat transfer decreases and the wall superheat becomes higher (see Fig. \ref{fig:dT} at H/D of 42.5 and 57.5). One cycle of bubble nucleation, transformation into a slug and its destruction is shown in the Fig. \ref{fig:Slug}.

The maximum measured enhancement of heat transfer was at $S=1.75$, superheat of 5.8 K and H/D of 27.5. In that conditions, the heat flux density imposed to the surface of the tube was 10 kW/m$^2$, that is two times higher than that for $S=10.75$ or $S=6.25$ (see Fig. \ref{fig:dT}). At the same time, the enhancement of HTC versus heat flux density was 20\%.

When the slug flow exists, the nucleation of bubbles occurs also preferably at the bottom part of the bundle. Thermal boundary layer destruction by slugs is significant, because they push liquid through the bundle improving HTC and reducing wall superheat below the value necessary for bubble nucleation. Furthermore, the evaporation of the liquid layer between the slugs and tubes walls contribute to HTC enhancement. Thus, the effect of slugs on the heat transfer is also associated with their life span, that is inversely proportional to heat flux density and also depends on the position along the tubes where the bubble was nucleated and slugs coalescence. The existence of the slug flow can be clearly noted by the change in the slope of the boiling curves shown in Fig. \ref{fig:dT}. 

%The measurements of the slug height versus time revealed that the slugs flow with the constant speed (see Fig. \ref{fig:Slug} (bottom).

\begin{figure}
\begin{minipage}[h]{0.18\linewidth}
\center{\includegraphics[width=1\linewidth]{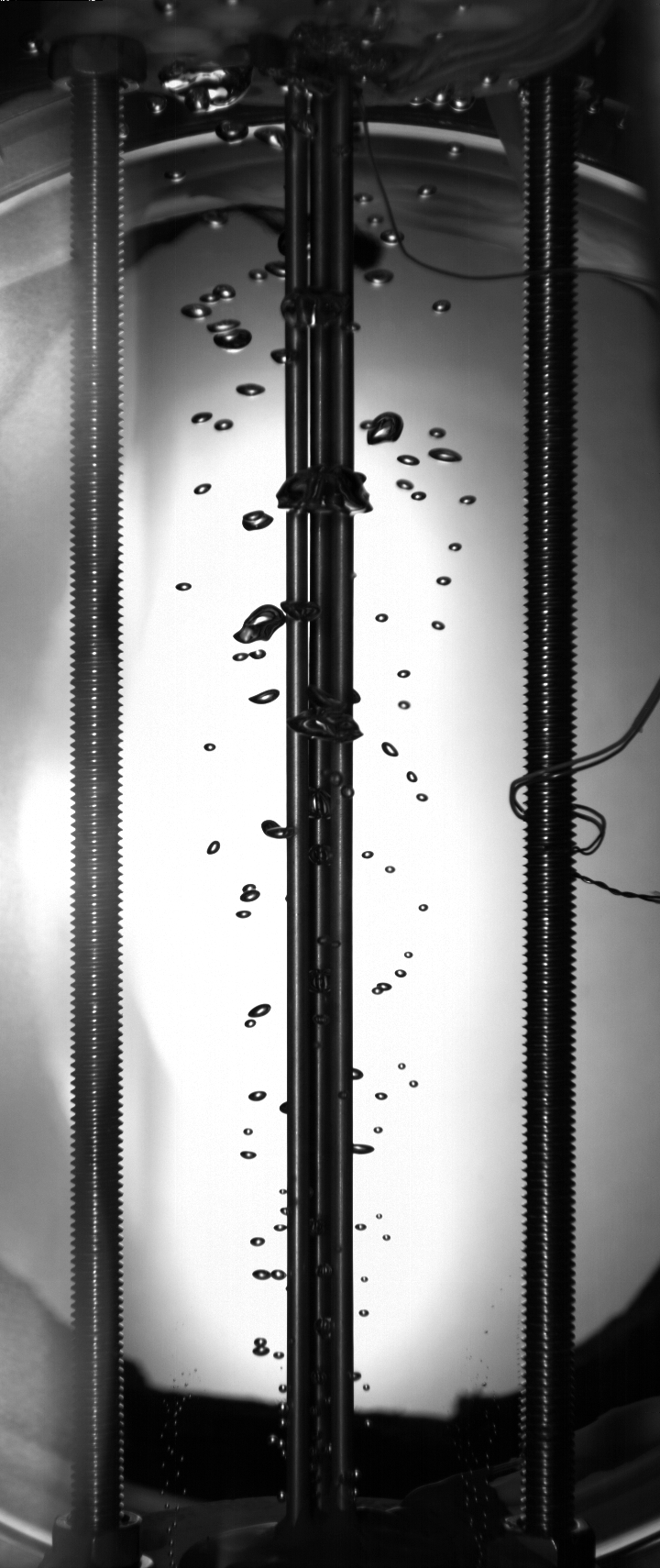}}  7.1 kW/m$^2$ \\
\end{minipage}
\begin{minipage}[h]{0.18\linewidth}
\center{\includegraphics[width=1\linewidth]{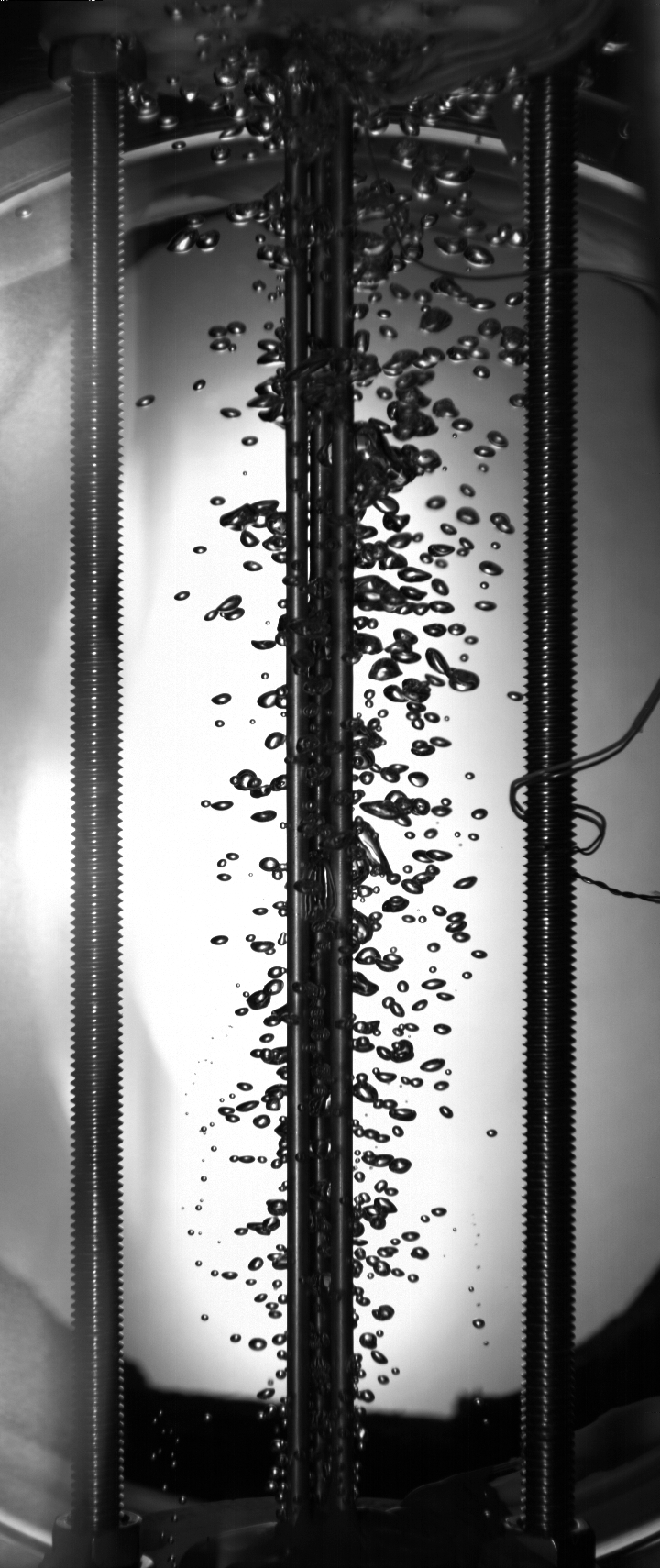}}  17.9 kW/m$^2$ \\
\end{minipage}
\begin{minipage}[h]{0.18\linewidth}
\center{\includegraphics[width=1\linewidth]{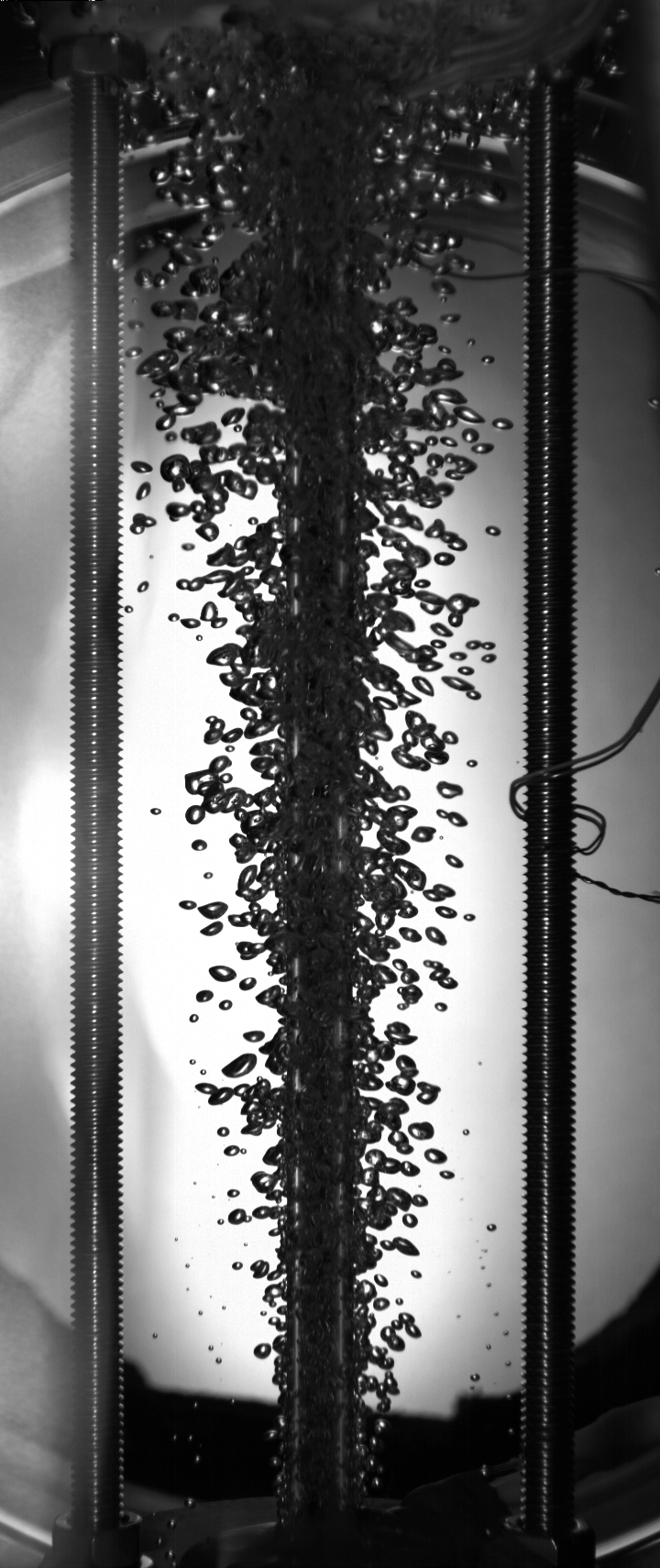}} 35.5 kW/m$^2$ \\
\end{minipage}
\begin{minipage}[h]{0.18\linewidth}
\center{\includegraphics[width=1\linewidth]{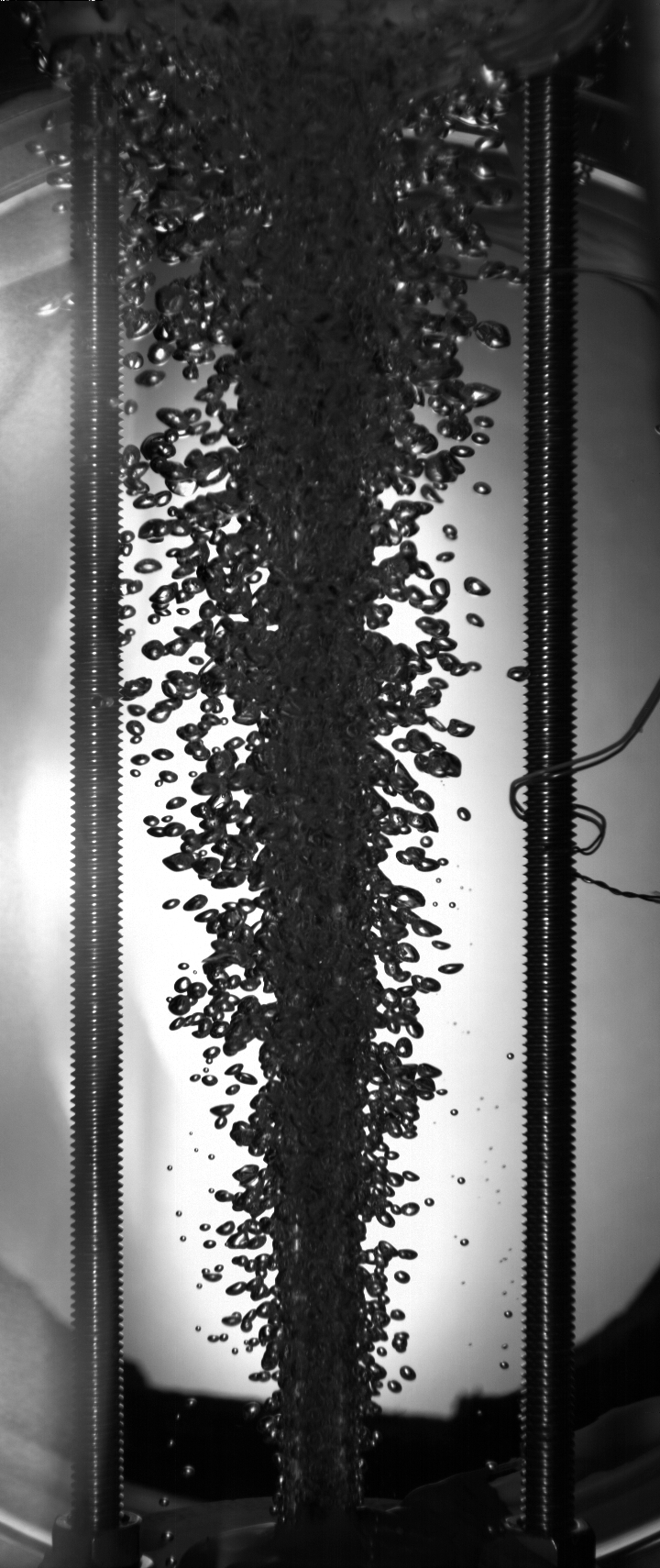}} 53.1 kW/m$^2$ \\
\end{minipage}
\begin{minipage}[h]{0.18\linewidth}
\center{\includegraphics[width=1\linewidth]{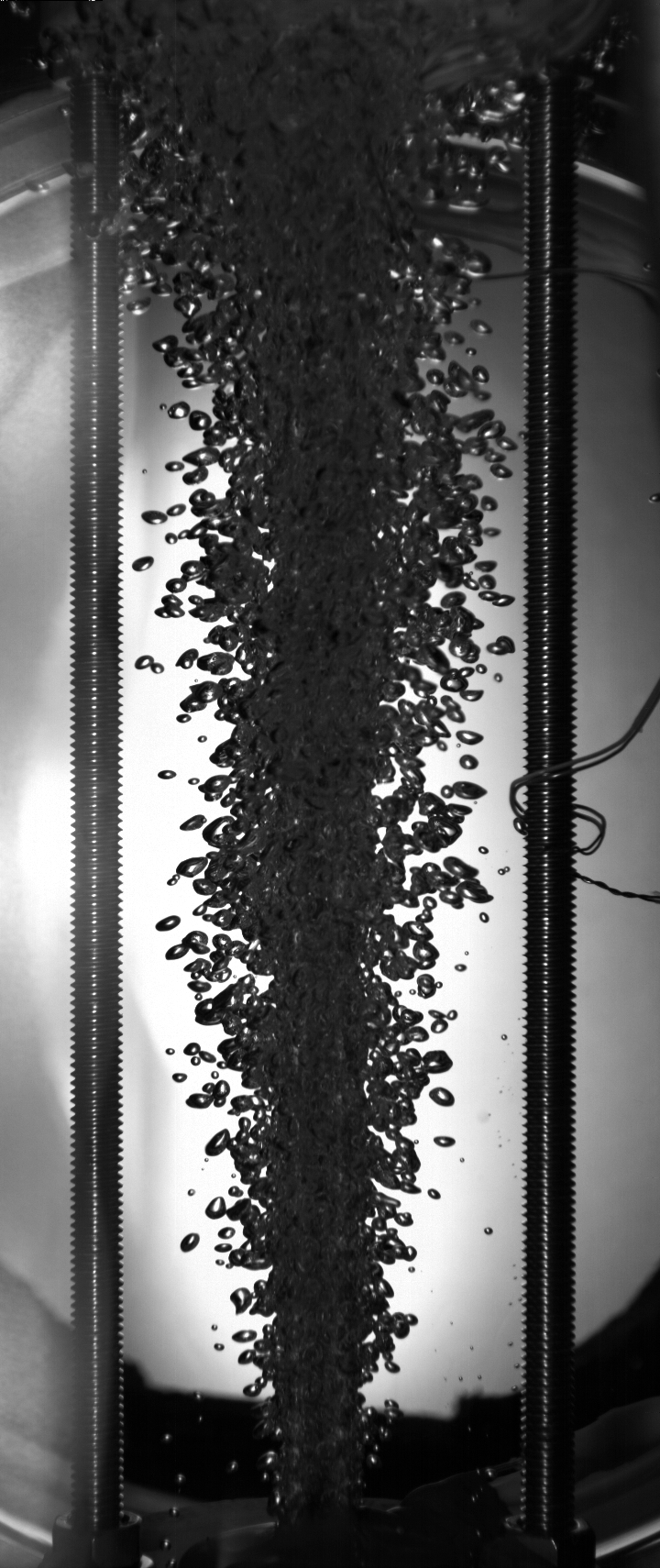}} 70.8 kW/m$^2$ \\
\end{minipage}
\vfill
\begin{minipage}[h]{0.18\linewidth}
\center{\includegraphics[width=1\linewidth]{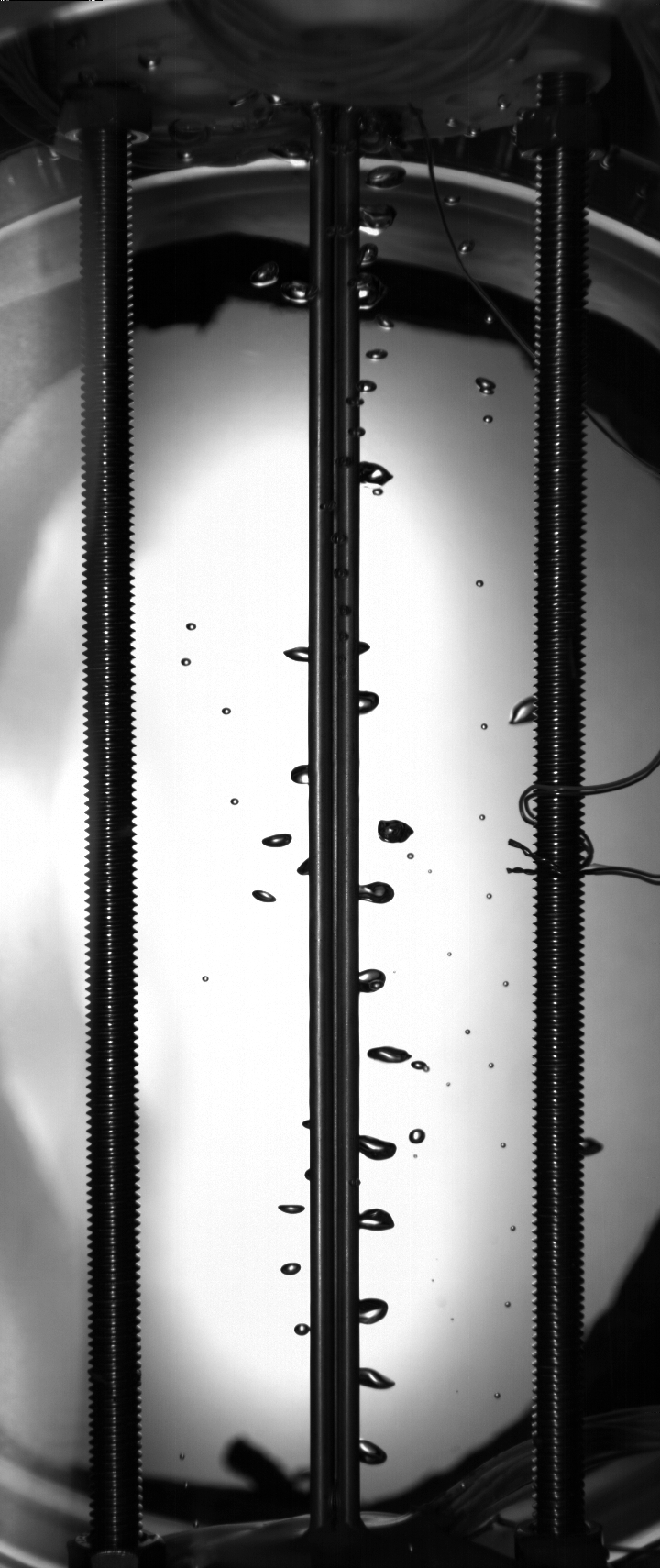}}   7.1 kW/m$^2$ \\
\end{minipage}
\begin{minipage}[h]{0.18\linewidth}
\center{\includegraphics[width=1\linewidth]{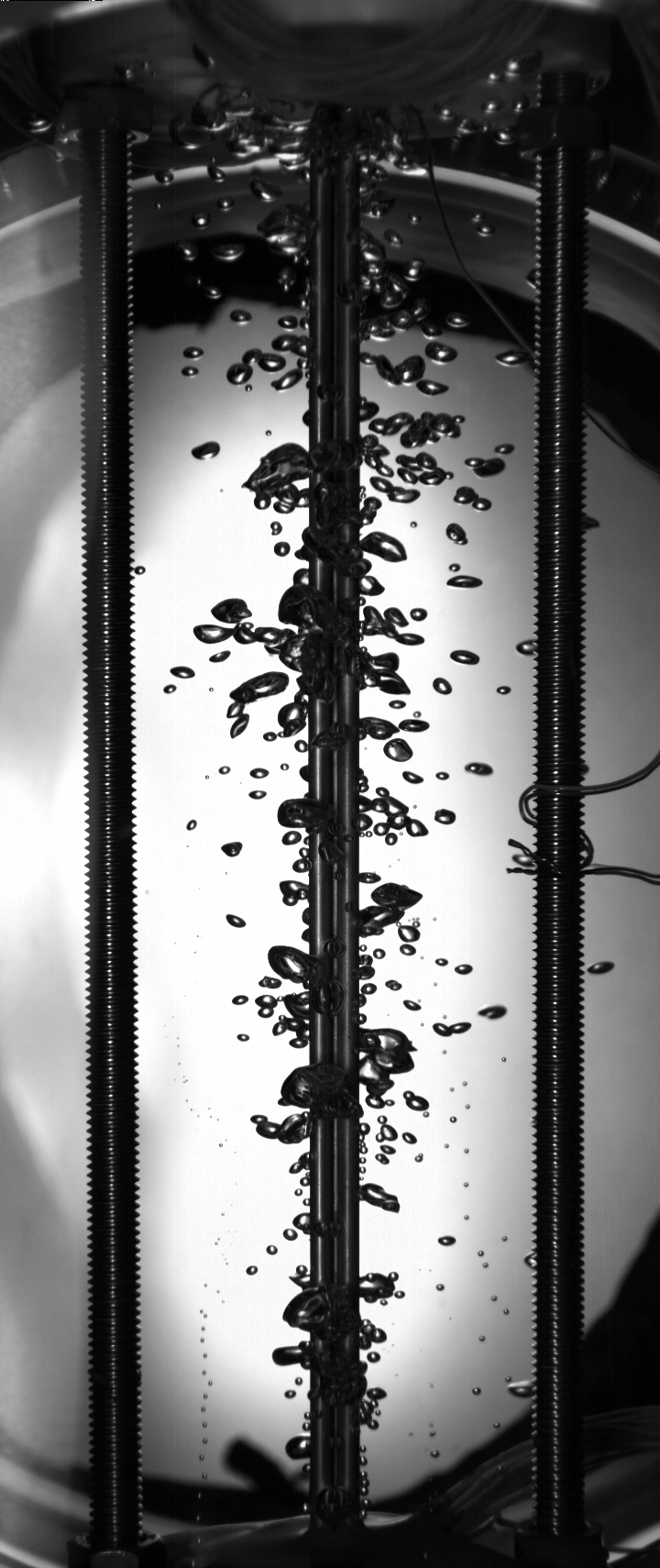}}   17.8 kW/m$^2$\\
\end{minipage}
\begin{minipage}[h]{0.18\linewidth}
\center{\includegraphics[width=1\linewidth]{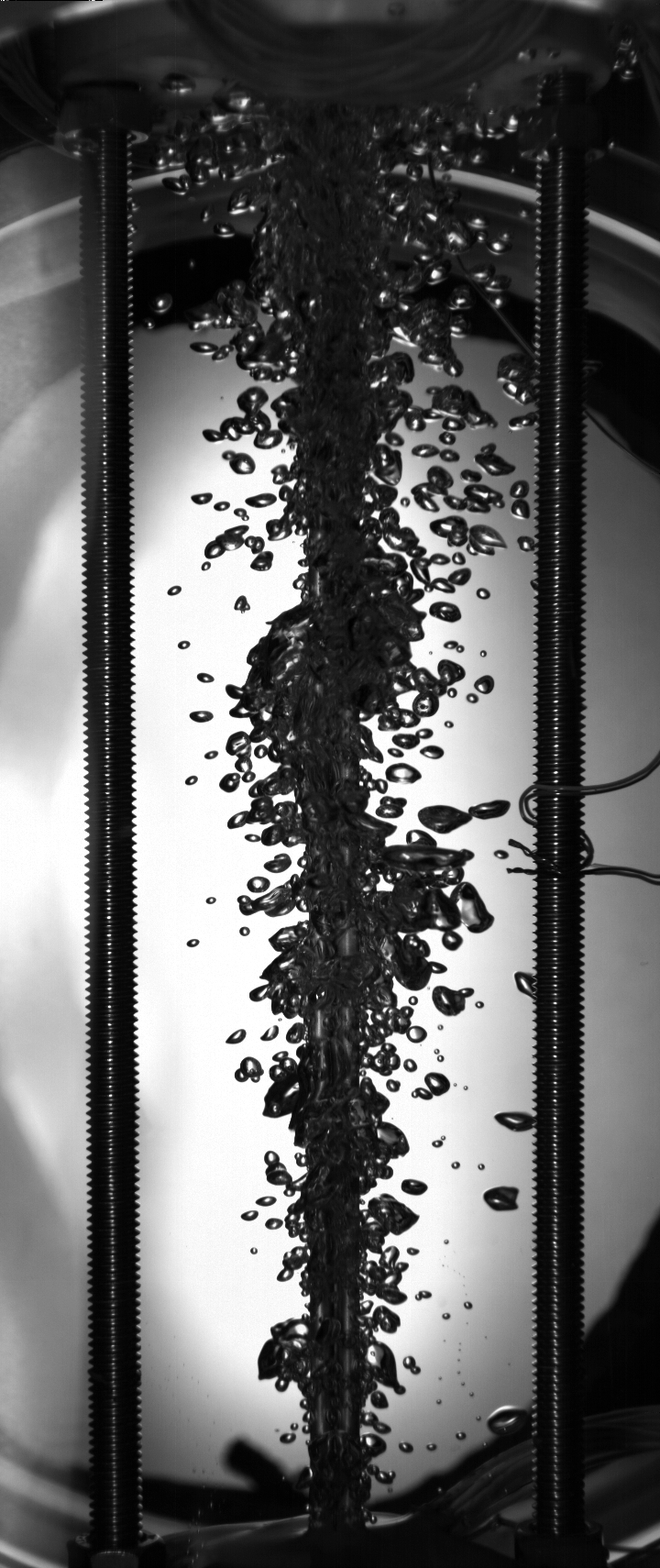}} 35.4 kW/m$^2$ \\
\end{minipage}
\begin{minipage}[h]{0.18\linewidth}
\center{\includegraphics[width=1\linewidth]{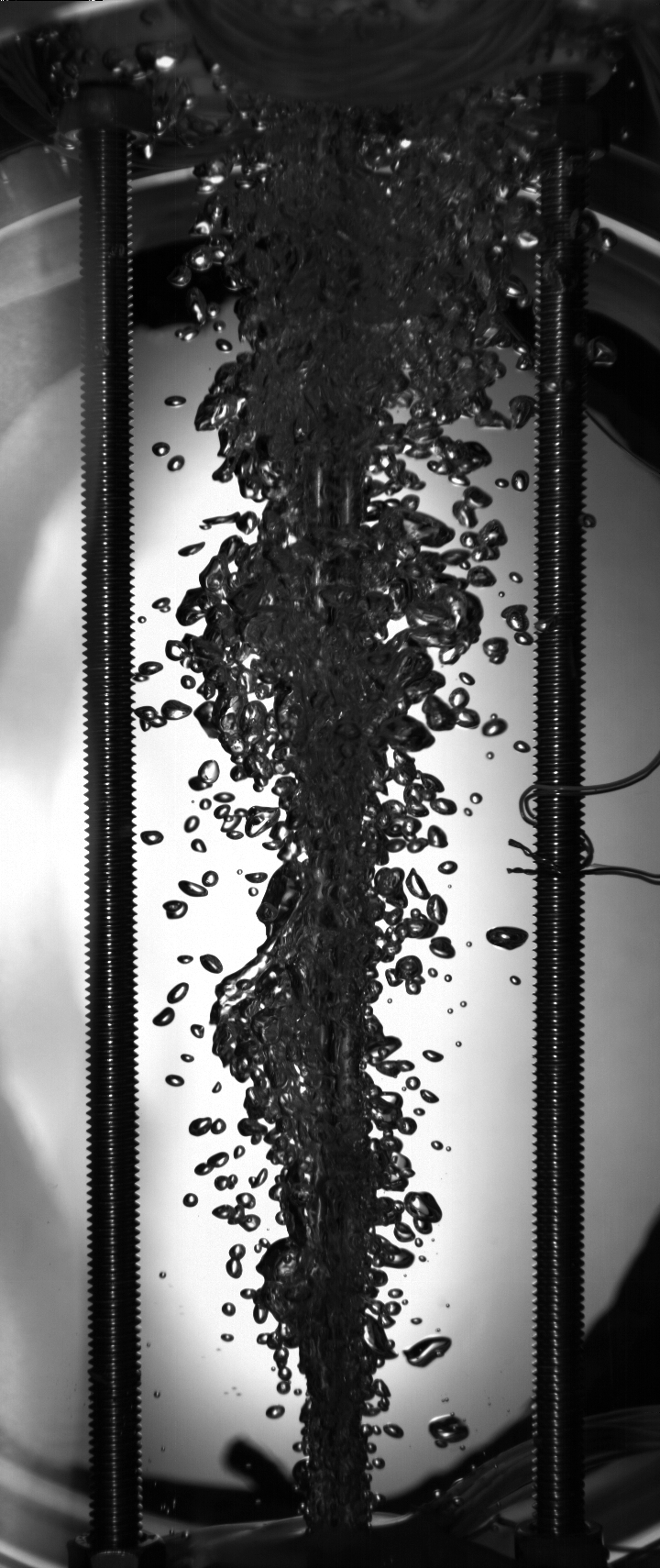}}  53.0 kW/m$^2$ \\
\end{minipage}
\begin{minipage}[h]{0.18\linewidth}
\center{\includegraphics[width=1\linewidth]{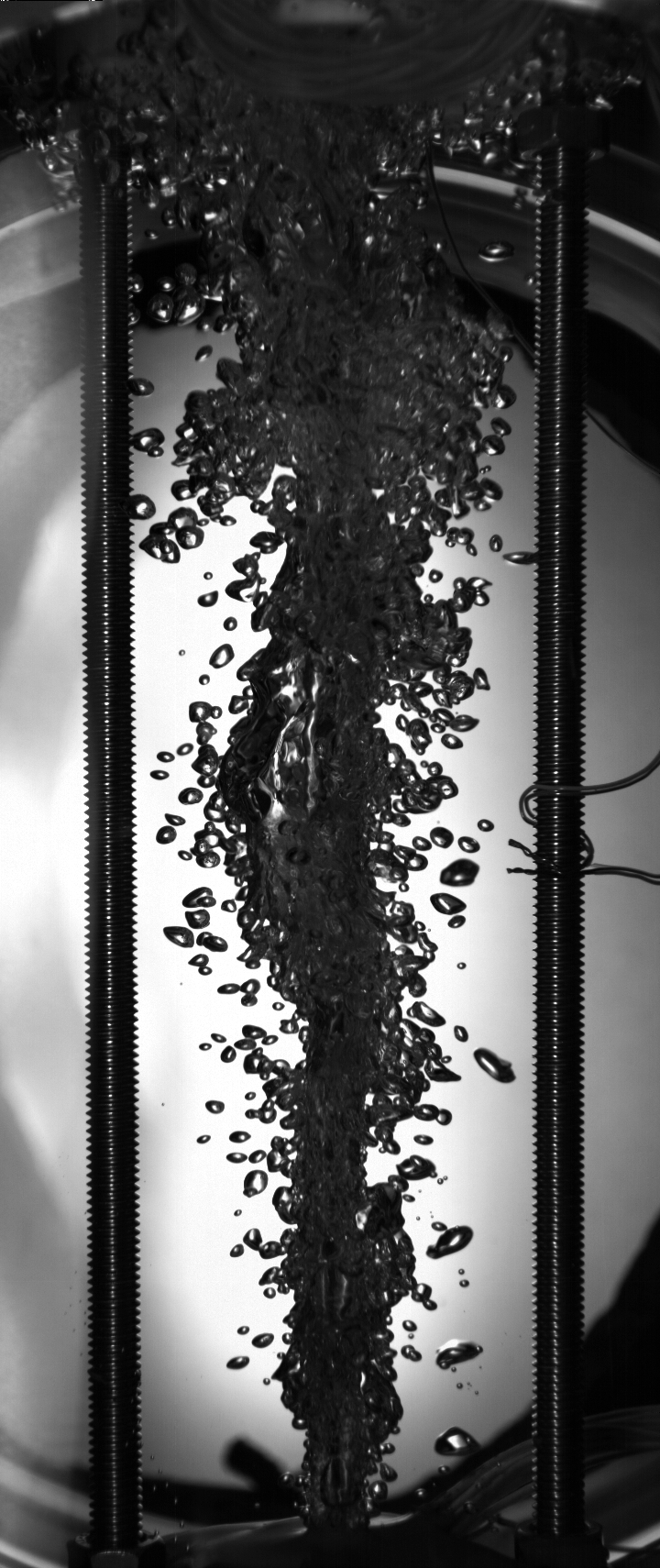}} 70.6 kW/m$^2$ \\
\end{minipage}
\caption{Snapshots of boiling process at different heat fluxes and spacings: Top row $S=1.75$; Bottom row $S=0.25$.}
\label{fig:SP1 and SP0}
\end{figure}

At $S=1.75$ mm the "flambeau" effect is more pronounced compared to $S=10.75$ and $S=6.25$ (see Fig. \ref{fig:SP1 and SP0} top row). Nonetheless, this effect was found again insignificant, because in the mean range of heat flux density 20-45 kW/m$^2$, the HTC at this spacing is similar to that of $S=10.75$ and $S=6.25$ mm. It is revealing that the process of bubble nucleation is dominant among others. However, at heat flux density higher than 45 kW/m$^2$ the bubbles tend to coalescence that reduces the vapor removing rate and consequently increases the wall superheat (Fig. \ref{fig:dT}).

At the shortest spacing of 0.25 mm the bubbles are significantly larger due to coalescence and higher volumetric energy density release in the gap between the tubes that locally increase the number of active nucleation sites and as a consequence increase the vapor generation rate. At such narrow spacing, slugs were only observed at a very low heat fluxes below 7 kW/m$^2$ at H/D of 12.5 (see Fig. \ref{fig:dT}). That mean that the slug flow existence is also defined by the spacing.

\begin{figure}
\center
\includegraphics[width=0.45\linewidth]{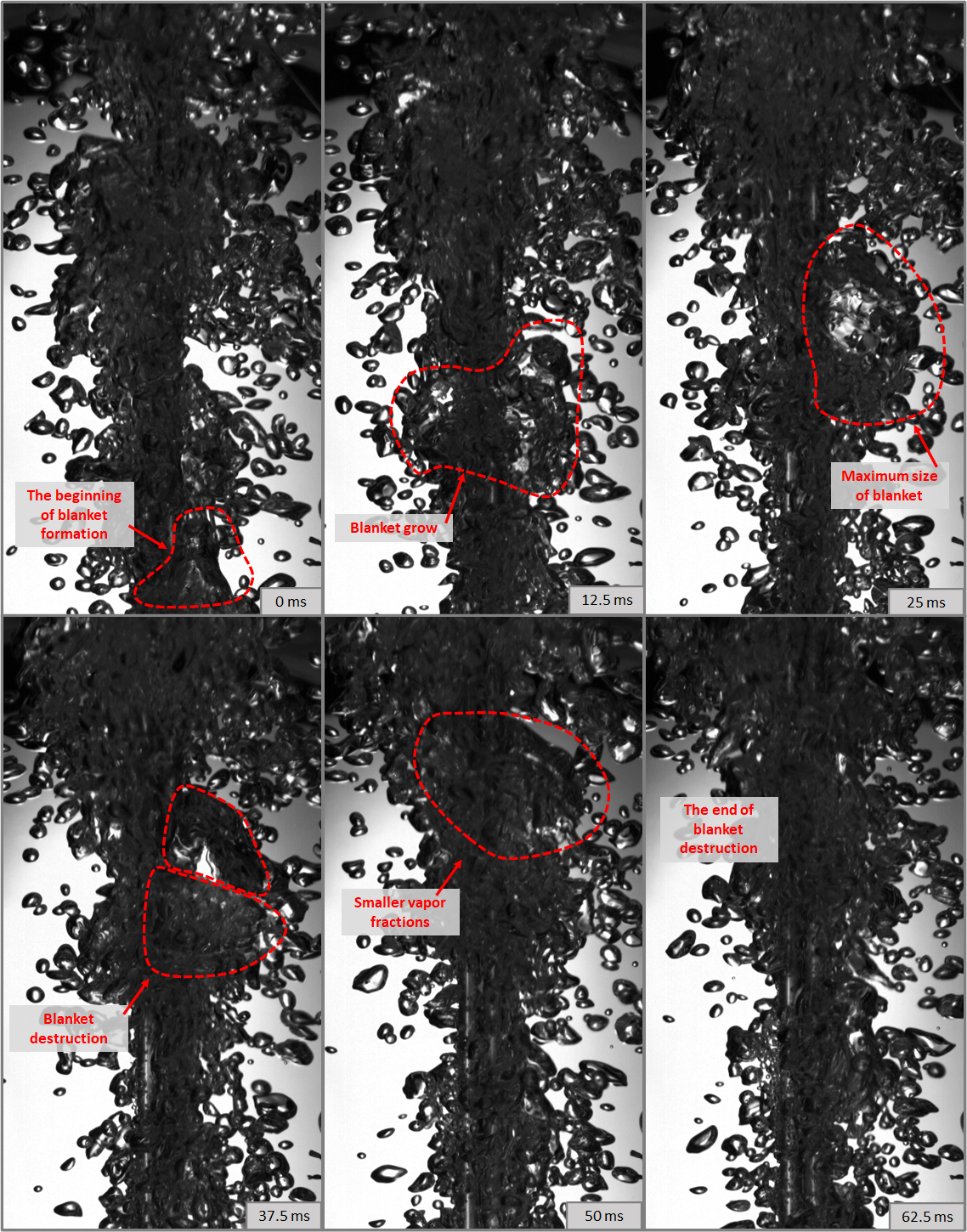}%
\caption{Snapshots of one blanket formation-destruction cycle at $q=53.0$ kW/m$^2$ and $S=0.25$ (see also movie 2 in supplementary material}
\label{fig:Blanket}
\end{figure}

In this configuration, the wall superheat is always lower at the bottom part of the bundle due to stronger turbulence produced by explosive-like grow of larger bubbles (see Fig. \ref{fig:dT} H/D=12.5 and Fig. \ref{fig:SP1 and SP0} bottom row). In spite of the differences in bubble behavior, at H/D=27.5 the wall superheat is comparable to those observed at $S=10.75$, $S=6.25$ and $S=1.75$. The reason is that at this level large bubbles start to create vapor blankets that first increase wall superheat along the height of the tubes and finally make it higher at H/D=42.5 (See Fig. \ref{fig:dT} and Fig. \ref{fig:SP1 and SP0} bottom row). However, the vapor blankets are not stable and get destroyed into smaller vapor fractions while moving upwards along the tubes. Consequently, at H/D=57.5 the recorded wall superheat is close to those measured at longer spacings up to about 50 kW/m$^2$, because all vapor blankets get destroyed before reaching this height of the tubes. A cycle of blanket formation, growth and destruction is shown in Fig. \ref{fig:Blanket}. When increasing the heat flux density above 50 kW/m$^2$ more and more vapor blankets reach the level of H/D=57.5 that start to increase local wall superheat.

\subsection{Interrelation of bubble departure diameter and spacing}
As known, the internal boiling characteristics play the main role in boiling process \cite{nikulin2018study,pioro2004nucleate,gerardi2011infrared,pontes2020effect}. In particular, bubble departure diameters were measured in order to clarify their interrelation with the tubes spacing and slug flow formation. For each spacing the high speed recordings were processed according to the procedure described in Section \ref{sec.BDD}. The data was obtained at fixed heat flux of 17.8 $\pm$ 0.1 kW/m$^2$. The obtained results are given in Table \ref{tabular:BDD}. It can be seen, that when the bubble departure diameter is much lower than the spacing $D_b \ll S$ there is no slugs in the tube bundle. However, when the bubble departure diameter is comparable to the spacing $D_b \sim S$ the slug flow exists in the bundle when the boiling is not fully developed. The spacings much shorter than bubble departure diameter $D_b \gg S$ provoke large vapor fractions creation that may improve heat transfer through the high turbulence or deteriorate it through the vapor blankets.

\begin{table}
\caption{Bubble departure diameters measurements at 17.8 $\pm$ 0.1 kW/m$^2$}
\label{tabular:BDD}
\begin{center}
\begin{tabular}{lllll}
\hline
Spacing number  & 1     &     2 &    3 &   4 \\ 
\hline
$S$, mm         & 10.75  &  6.25 & 1.75 & 0.25  \\
\hline
Number of nucleation cites analyzed      & 8  &  7 & 1 & 2  \\
Average $D_b$, mm        & 1.2  &  1.5 & 1.2 & 1.3    \\
\hline
\end{tabular}
\end{center}
\end{table}

\section{Conclusions}

Boiling process of a simplified tubes bundle that consist of three vertical tubes has been studied systematically and the following conclusions are drawn:

-Long spacings (6.25 and 10.75 mm) that are much longer than $D_b$ (1.2-1.5 mm) are not influencing the boiling process;

-Short spacings (1.75 mm) on the order of $D_b$ promote slug flow in the bundle, that may increase the heat transfer coefficient up to two times versus wall superheat and up to 20\% versus heat flux density;

-Spacings that are much shorter than $D_b$ (0.25 mm) are very beneficial for the HTC of tubes bundles with low length to diameter ratios (up to L/D=25). However, at higher L/D vapor blankets deteriorate heat transfer.

From the performed study it is clear, that for the M-TES system utilising boiling to discharge cylindrically shaped and vertically oriented PCM capsules, the spacings much longer than $D_b$ are not efficient for both high discharging rate and storage capacity of the system. Spacings on the order of $D_b$ are better solution for the systems with high L/D ratio of PCM capsules that are working at low heat fluxes. Allocation of vertical arrays of PCM capsules with narrow spacings and low L/D ratio is probably the best solution for M-TES systems with both high storage capacity and discharge rate that are capable to work in a wide range of heat fluxes.

\section{Supplementary material}
Movie 1 (\url{https://drive.google.com/file/d/1w4dzG_0dQRD7qj6b4B6vlsqUNnbC8rd0/view?usp=sharing})

Movie 2 (\url{https://drive.google.com/file/d/1hje9U1L9kFh0H2ghqmQuuFbmcj_bUO5E/view?usp=sharing})

\section*{Acknowledgements}
The authors are grateful for the financial support from FSWEET-TES project (RTI2018-099557-B-C21), funded by FEDER/Ministerio de Ciencia e Innovación – Agencia Estatal de Investigación and Elkartek CICe2020 project (KK-2020/00078) funded by Basque Government.

\newpage

%(top) and its position versus time (bottom)

\afterpage{

\bibliographystyle{elsarticle-num-names} %In sequence

\bibliography{references}
}

\end{document}